%% file: vine_knockoffs.tex
\documentclass[11pt,a4paper]{article}

\usepackage[top=3cm, bottom=3cm, left=2.5cm, right=2.5cm]{geometry}

\usepackage[ruled,vlined]{algorithm2e}

\usepackage[english]{babel}
\usepackage{amsmath}
\usepackage{amsfonts}
\usepackage{amssymb}
\usepackage{mathtools}
\usepackage{dsfont}

\usepackage{graphicx}
\usepackage{subcaption}
\usepackage{placeins}

\newcommand{\ps}{;\,}

\DeclareFontFamily{U}{mathx}{}
\DeclareFontShape{U}{mathx}{m}{n}{<-> mathx10}{}
\DeclareSymbolFont{mathx}{U}{mathx}{m}{n}
\DeclareMathAccent{\widehat}{0}{mathx}{"70}
\DeclareMathAccent{\widecheck}{0}{mathx}{"71}
\DeclareMathAccent{\widetilde}{0}{mathx}{"72}

\allowdisplaybreaks

\usepackage[dvipsnames]{xcolor}
\usepackage{todonotes}

\usepackage[backend=bibtex8,style=authoryear-comp,maxnames=2,minnames=1,maxbibnames=99,doi=false,url=false,eprint=false,isbn=false,firstinits=true]{biblatex}

\usepackage[hidelinks]{hyperref}

\addto\extrasenglish{%

}

\bibliography{vine_ko.bib}

\numberwithin{equation}{section}

\usepackage{authblk}

\author{Malte S. Kurz\thanks{malte.kurz@tum.de}}
\affil{TUM School of Management\protect\\
Technical University of Munich\protect\\
Arcisstr.\ 21, 80333 Munich, Germany}

\title{Vine copula based knockoff generation for high-dimensional controlled variable selection}

\begin{document}

\maketitle

\input{abstract}

\input{intro}

\input{model_x_ko}

\input{gaussian_ko}

\input{gaussian_cop_ko}

\input{vine_ko}

\input{sim}

\input{outlook}

%

\section*{Acknowledgements}
We acknowledge funding by the Deutsche Forschungsgemeinschaft (DFG, German Research Foundation) -- Project Number 431701914.

\newpage
\printbibliography

\newpage
\appendix
\input{algos}

\end{document}

%% file: abstract.tex
\begin{abstract}
Vine copulas are a flexible tool for high-dimensional dependence modeling.
In this article, we discuss the generation of approximate model-X knockoffs with vine copulas.
It is shown how Gaussian knockoffs can be generalized to Gaussian copula knockoffs.
A convenient way to parametrize Gaussian copulas are partial correlation vines. 
We discuss how completion problems for partial correlation vines are related to Gaussian knockoffs.
A natural generalization of partial correlation vines are vine copulas which are well suited for the generation of approximate model-X knockoffs.
We discuss a specific D-vine structure which is advantageous to obtain vine copula knockoff models.
In a simulation study, we demonstrate that vine copula knockoff models are effective and powerful for high-dimensional controlled variable selection.
\end{abstract}

%% file: intro.tex
\section{Introduction}
In various fields, like economics, finance, biology or medicine,
researchers and practitioners try to identify important variables explaining a response variable of interest.
The set of potential explanatory variables is often high-dimensional.
Therefore, appropriate statistical tools are necessary to control the false discovery rate.
Model-X knockoffs \parencite{candes2018} can be used in such situations for high-dimensional controlled variable selection.

Knockoffs have been introduced by \textcite{barber2015} and been generalized to model-X knockoffs in \textcite{candes2018}.
The general idea of knockoffs is the following:
Denote the response variable by $Y$ and the $d$-dimensional vector of explanatory variables by $X_{1:d}$.
For the explanatory variables $X_{1:d}$, knockoff copies $\widetilde{X}_{1:d}$ are constructed.
To obtain valid knockoffs, two properties need to be satisfied.
First, it is required that conditionally on the true explanatory variables $X_{1:d}$, the knockoffs $\widetilde{X}_{1:d}$ are not associated with the response variable $Y$.
At the same time, the knockoff copies $\widetilde{X}_{1:d}$ need to be constructed in a way that their distributional structure is very similar to the original variables $X_{1:d}$.\footnote{%
A formal definition of knockoffs will be given in \autoref{sec_model_x_ko}.}
In variable selection procedures, these knockoffs are then added as control variables.
Only if an original variable $X_j$ is particularly more relevant than its knockoff counterpart $\widetilde{X}_j$, it will be considered as important explanatory variable for the response $Y$.
The knockoffs framework by \textcite{barber2015} and \textcite{candes2018} is then constructed in a way such that the false discovery rate is, at least approximately, controlled.

A key part of the knockoffs framework for high-dimensional controlled variable selection is the knockoff generation procedure or model.
Different methods have been proposed and analyzed in the literature.
The simplest approach relies on a multivariate Gaussian distribution which is obtained by matching the first and second moments \parencite{candes2018}.
More flexible alternatives are based on hidden Markov models \parencite{sesia2018} or variational auto-encoders \parencite{liu2018}.
Deep learning methods for knockoff generation have been analyzed in \textcite{romano2020,sudarshan2020} and generative adversarial networks in \textcite{jordon2018}.

In the following, we will propose a knockoff generation method based on vine copulas.
Vine copulas \parencite{aas2009,bedford2001,joe1997} are a flexible model class for high-dimensional dependence modeling and have already been applied as high-dimensional generative models \parencite[see][]{tagasovska2019}.
We will discuss how Gaussian knockoffs can be generalized to Gaussian copula knockoffs.
Gaussian copula knockoffs in particular allow for more flexible marginal distributions, i.e., non-normal marginal distributions.
Gaussian copulas can be parametrized with so-called partial correlation vines \parencite{bedford2002,kurowicka2003}.
We will explain how completion problems for partial correlation vines are related to the construction of knockoffs.
Partial correlation vines can be generalized to vine copulas in order to allow for more flexible dependence structures.
We will introduced a specific D-vine copula model which is particularly well suited for constructing approximate model-X knockoffs.
In a simulation study it is demonstrated that vine copula based knockoffs are effective and powerful for high-dimensional controlled variable selection.
An implementation of all three knockoff methods (Gaussian knockoffs, Gaussian copula knockoffs and vine copula knockoffs) is available in an accompanying \texttt{Python} package \texttt{vineknockoffs} \parencite{kurz2022}.

The paper is structured as follows.
In \autoref{sec_model_x_ko}, we will repeat the most important concepts of the model-X knockoff framework.
Gaussian knockoffs are discussed in \autoref{sec_gau_ko}.
The generalization to Gaussian copula knockoffs, partial correlation vines and vine copulas are discussed in \autoref{sec_gaucop_ko}.
In \autoref{sec_vine_ko}, we introduce vine copula based knockoffs and discuss implementation details.
A simulation study, where the finite sample performance for high-dimensional controlled variable selection is analyzed, is presented in \autoref{sec_sim}.
Concluding remarks are given in \autoref{sec_concl}.

%% file: model_x_ko.tex
\section{Model-X knockoffs}\label{sec_model_x_ko}
A random vector $\widetilde{X}_{1:d} \in \mathbb{R}^d$ is called a model-X knockoff copy of $X_{1:d} \in \mathbb{R}^d$, if the following properties are satisfied
\begin{align}
(X_{1:d}, \widetilde{X}_{1:d}) &\stackrel{d}{=} (X_{1:d}, \widetilde{X}_{1:d})_{\text{swap}(\mathcal{S})},
\qquad \text{for each }  \mathcal{S}  \subseteq 1{:}d := \lbrace 1, \ldots, d \rbrace, \label{eq_ko_1} \\
Y &\perp \widetilde{X}_{1:d} \;|\; X_{1:d}, \label{eq_ko_2} 
\end{align}
where $\stackrel{d}{=}$ denotes equality in distribution.\footnote{
Originally, knockoffs have been introduced by \textcite{barber2015} under the assumption that the covariates are fixed.
The term model-X knockoffs was introduced by \textcite{candes2018} who treat the covariates $X_{1:d}$ as random variables to be able to apply knockoffs for high-dimensional settings.}

The first knockoff property \eqref{eq_ko_1} implies that the joint distribution of the vector of covariates $X_{1:d}$ together with the vector of its knockoffs $\widetilde{X}_{1:d}$ is invariant against any kind of swap.
A swap with subset $\mathcal{S}  \subseteq 1{:}d$ is obtained by swapping the entries $X_j$ and $\widetilde{X}_{j}$ for each $j \in \mathcal{S}$ in the augmented vector $(X_{1:d}, \widetilde{X}_{1:d})$.
For example for $d=3$, we can consider the subset $\mathcal{S} = \lbrace 1, 3 \rbrace$ and the knockoff property \eqref{eq_ko_1} becomes
\begin{align*}
\left(X_1, X_2, X_3, \widetilde{X}_1, \widetilde{X}_2, \widetilde{X}_3\right)
\stackrel{d}{=} \left(\widetilde{X}_1, X_2, \widetilde{X}_3, X_1, \widetilde{X}_2, X_3\right).
\end{align*}
The second knockoff property \eqref{eq_ko_2} is satisfied if, given the original explanatory variables $X_{1:d}$, the knockoffs $\widetilde{X}_{1:d}$ have no effect on the response variable $Y$.
Finding knockoff generation methods such that the first knockoff property \eqref{eq_ko_1} is satisfied can be challenging.
For some specific distributions, like the multivariate normal distribution, it is possible to obtain exact knockoff copies \parencite[see for example][]{candes2018,sesia2018,gimenez2019}.
If the distribution of $X_{1:d}$ is more complex, various methods have been proposed in the literature.
These methods can be used to construct knockoffs that to a certain extend approximately satisfy the knockoff property \eqref{eq_ko_1}.
In contrast, the second knockoff property \eqref{eq_ko_2} is easily satisfied, if the outcome variable $Y$ is not used to construct the knockoffs.

To repeat some key terms for controlled variable selection and model-X knockoffs, we consider the following problem.
Assume that we have obtained a sample from a response variable of interest $Y$ together with covariates $X_{1:d}$ which might explain $Y$.
The goal is to identify a subset of $X_{1:d}$ containing important variables which have an effect on $Y$.
To formalize this, lets assume that the response only depends on a (small) subset of variables $\mathcal{S} \subset \lbrace 1, \ldots, d \rbrace$ such that, conditionally on $\lbrace X_i \rbrace_{i \in \mathcal{S}}$, the outcome variable $Y$ is independent of all other covariates.
We further denote by $\widehat{\mathcal{S}}$ the set of important variables which has been identified with a variable selection procedure.
Usually, such variable selection procedures are designed in a way that the false discovery rate is controlled, i.e.,
\begin{align*}
\mathbb{E}\left[
\frac{\# \lbrace i: i\in \widehat{\mathcal{S}} \setminus \mathcal{S} \rbrace}{\# \lbrace i: i\in \widehat{\mathcal{S}}\rbrace}
\right]
\leq q,
\end{align*}
for some nominal level $q \in (0,1)$ and with the convention $\frac{0}{0}=0$.

It has been shown in \textcite{candes2018} that model-X knockoffs are a variable selection method where the false discovery rate is controlled.
In the following, we will briefly repeat the most important steps of the model-X knockoffs framework.
A first key element is a method to construct model-X knockoffs satisfying the knockoff properties \eqref{eq_ko_1} and \eqref{eq_ko_2}.
Additionally, some measures of feature importance $Z_i$ and $\widetilde{Z}_i$ are required for each variable $X_i$, $1 \leq i \leq d$, and their knockoff copies $\widetilde{X}_{i}$, $1 \leq i \leq d$, respectively.
These measures of feature importance can be obtained from standard ML-methods like for example a Lasso or elastic net regression of $Y$ on the augmented vector $(X_{1:d}, \widetilde{X}_{1:d})$.
The feature importance scores of each variable and its knockoff are then combined to a knockoff statistic, e.g., $W_i = Z_i - \widetilde{Z}_i$.
This knockoff statistic is antisymmetric, so that a large positive value of the knockoff statistic $W_i$ is an indication for an important variable $X_i$.
At the same time for an unimportant variable $X_i$, positive and negative values should be equally likely for the knockoff statistic $W_i$.
The estimated set of important variables with the model-X knockoffs framework, while controlling the false discovery rate, is then obtain as 
$\widehat{\mathcal{S}} := \lbrace i: W_i \geq \tau_q \rbrace$.
Here, the threshold $\tau_q$ is given by \parencite{barber2015,candes2018}\footnote{
Note that recently proposed extensions can be employed to derandomize knockoffs and / or find better thresholds for the knockoff filter \parencite[see for example][]{ren2021,luo2022,emery2019,gimenez2019b}.
Many of these methods try to improve the stability of knockoff filters by generating multiple or simultaneous knockoffs and combining them in an appropriate way.
These advanced methods or extensions could also be combined with the vine copula knockoff generation method.
For the sake of simplicity, this is left for future research.}
\begin{align*}
\tau_q = \min \left\lbrace t >0: \frac{1+ \# \lbrace i: W_i \leq -t \rbrace}{\# \lbrace i: W_i \geq t \rbrace} \leq q \right\rbrace.
\end{align*}
The validity and quality of model-X knockoffs fundamentally depends on the procedure used for generating knockoffs which satisfy the properties \eqref{eq_ko_1} and \eqref{eq_ko_2}.
In the following, we will propose a new such knockoff generation method.
The new method utilizes vine copulas which are a powerful tool for high-dimensional dependence modeling.

%% file: gaussian_ko.tex
\section{Gaussian knockoffs}\label{sec_gau_ko}
Assume that the $d$-dimensional covariates $X_{1:d} \in \mathbb{R}^d$ are multivariate normally distributed, i.e., $X_{1:d} \sim \mathcal{N}_d(0, \Sigma)$, where $\mathcal{N}_d(\mu, \Sigma)$ denotes a $d$-dimensional normal distribution with expectation $\mu$ and covariance matrix $\Sigma$.
Model-$X$ Knockoffs $\widetilde{X}_{1:d} \in \mathbb{R}^d$ can then be obtained from the following joint normal distribution \parencite{candes2018}
\begin{align}
(X_{1:d}, \widetilde{X}_{1:d}) \sim \mathcal{N}_{2d}(0, G),
\qquad\text{ where }\qquad
G = \left(\begin{matrix}
\Sigma & \Sigma - \text{diag}(s) \\
\Sigma - \text{diag}(s) & \Sigma
\end{matrix}\right)
\label{eq_gau_ko}%
\end{align}
and $\text{diag}(s)$ is a diagonal matrix such that $G$ is positive semidefinite.
Typically, the vector $s$ is obtained by solving a semidefinite program, see \textcite{candes2018}.

Having specified, or estimated, $\Sigma$, $s$ and therefore also $G$, knockoffs $\widetilde{X}_{1:d}$ can be simply obtained by sampling from the conditional distribution \parencite{candes2018}
\begin{align*}
\widetilde{X}_{1:d}|X_{1:d}=x_{1:d} \sim \mathcal{N}_{d}(\mu, V),
\end{align*}
with
\begin{align*}
\mu &= x_{1:d} - x_{1:d} \Sigma^{-1} \text{diag}(s), \\
V &= 2 \text{diag}(s) - \text{diag}(s) \Sigma^{-1} \text{diag}(s).
\end{align*}
This procedure to generate knockoffs will in the following be called \textit{Gaussian knockoffs}.
It is restricted to the case of multivariate normally distributed covariates $X_{1:d}$.
However, it can also be applied for covariates that are not normally distributed.
In such cases their distribution is approximated by a multivariate normal distribution and the knockoff generation procedure is sometimes also called \textit{second-order knockoffs} \parencite{candes2018}.\footnote{
We will in the following always use the term \emph{Gaussian knockoffs} irrespectively whether $X_{1:d}$ is multivariate normally distributed or not.}
The name reflects the fact that the procedure is designed in a way that first and second moments of the data are matched but not necessarily the entire distribution.

%% file: gaussian_cop_ko.tex
\section{Gaussian copula knockoffs, partial correlation vines and vine copulas}\label{sec_gaucop_ko}
The \textit{Gaussian knockoff} generation procedure is restricted to the multivariate normal case, i.e., it requires the distributional assumption $X_{1:d} \sim \mathcal{N}_d(0, \Sigma)$.
If this assumption fails to hold, Gaussian knockoffs might still be used.
However, depending on how well the true distribution of $X_{1:d}$ can be approximated with a multivariate normal distribution, Gaussian knockoffs might not produce valid results.

\subsection{Gaussian copula knockoffs}\label{sec_gaucop_gaucop_ko}
A straightforward generalization of Gaussian knockoffs is given by assuming a Gaussian copula for $X_{1:d}$, i.e.,
\begin{align*}
U_{1:d} := \big(F_1(X_1), \ldots, F_d(X_d)\big) \sim C_{d}^{Gau}(R),
\end{align*}
where $C_{d}^{Gau}(R)$ denotes a $d$-dimensional Gaussian copula with correlation matrix $R$ and
$F_1, \ldots, F_d$ are arbitrary absolutely continuous marginal distributions such that $X_1 \sim F_1, \ldots, X_d \sim F_d$.
Note that by definition
\begin{align*}
U_{1:d} \sim C_{d}^{Gau}(R)
\qquad \Leftrightarrow \qquad
Y_{1:d}:=\big(\Phi^{-1}(U_1), \ldots, \Phi^{-1}(U_d) \big) \sim \mathcal{N}_d(0, R),
\end{align*}
where $\Phi(\cdot)$ denotes the cdf of the univariate standard normal distribution.

To obtain valid knockoffs, we use the following model
\begin{align}
(U_{1:d}, \widetilde{U}_{1:d}) \sim C_{2d}^{Gau}(H),
\qquad\text{ where }\qquad
H = \left(\begin{matrix}
R & R - \text{diag}(r) \\
R - \text{diag}(r) & R
\end{matrix}\right).
\label{eq_gau_cop_ko}%
\end{align}
The vector $r$, depending on $R$, can be obtained as solution to the same kind of optimization problem being solved to obtain $s$ depending on $\Sigma$ (see \eqref{eq_gau_ko}).
We further set $\widetilde{X}_{1:d} := (F_1^{-1}(\widetilde{U}_1), \ldots, F_d^{-1}(\widetilde{U}_d))$ such that $\widetilde{X}_1 \sim F_1, \ldots, \widetilde{X}_d \sim F_d$.

Note that Gaussian knockoffs are a special case of Gaussian copula knockoffs which are obtained by setting all marginal distributions $F_1, \ldots, F_d$ to cdfs of the normal distributions $\mathcal{N}(0, \sigma_1^2)$, $\ldots$, $\mathcal{N}(0, \sigma_d^2)$ with the respective variances $\sigma_1^2, \ldots, \sigma_d^2$.

\input{pcorr_vines}
\input{vine_cop}

\subsection{Generating Gaussian copula knockoffs}
The procedure to generate Gaussian copula knockoffs can be summarized as follows:
First we transform $X_{1:d}$ to standard uniformly distributed random variables $U_{1:d} := \big(F_1(X_1), \ldots, F_d(X_d)\big)$.
Then we compute the PITs $W_{1:d} := \big( U_{1}, U_{2|1}, U_{3|1:2}, \ldots, U_{d|1:d-1} \big)$ from $U_{1:d}$ using a $d$-dimensional partial correlation D-vine (\autoref{algo_pits_dvine}).
Next, we sample $\widetilde{W}_{1:d} \sim \mathcal{U}([0, 1]^d)$ and apply the inverse probability integral transform (\autoref{algo_sim_dvine}) to $(W_{1:d}, \widetilde{W}_{1:d})$ with the $2d$-dimensional Gaussian copula $C_{2d}^{Gau}(H)$.
The corresponding $2d$-dimensional partial correlation D-vine forms the basis to obtain the vine copula parameters in order to apply \autoref{algo_sim_dvine}.
As a result we obtain $(U_{1:d}, \widetilde{U}_{1,d})$ from the conditional distribution $F_{\widetilde{U}_{1:d}|U_{1:d}}$.
Finally, we set $\widetilde{X}_{1:d} := (F_1^{-1}(\widetilde{U}_1), \ldots, F_d^{-1}(\widetilde{U}_d))$
in order to obtain knockoffs $\widetilde{X}_{1:d}$ as a sample from the conditional distribution $F_{\widetilde{X}_{1:d}|X_{1:d}}$.

If $(U_1,U_2) \sim C_2^{Gau}$, conditional distributions of $U_1$ given $U_2$, or $U_2$ given $U_1$, can be obtained via partial derivatives
\begin{align*}
F_{U_1|U_2}(u_1|u_2) &= \partial_2 C_2^{Gau}(u_1, u_2; \rho) = \Phi\left(\frac{\Phi^{-1}(u_1) - \rho \Phi^{-1}(u_2)}{\sqrt{1 - \rho}} \right), \\
F_{U_2|U_1}(u_2|u_1) &= \partial_1 C_2^{Gau}(u_1, u_2; \rho) = \Phi\left(\frac{\Phi^{-1}(u_2) - \rho \Phi^{-1}(u_1)}{\sqrt{1 - \rho}} \right).
\end{align*}
\autoref{algo_pits_dvine} can now be used to compute the PITs
$W_{1:d} := \big( U_1, U_{2|1}, U_{3|1:2}, \ldots, U_{d|1:d-1}\big)$
for a $d$-dimensional Gaussian copula $C_{d}^{Gau}(R)$.
The parameters of the bivariate Gaussian copulas are given by the partial correlation vine corresponding to the correlation matrix $R$ and the chosen D-vine structure.

A random sample from a Gaussian copula can be obtained by using inverse probability integral transforms.
For this, we define the inverse of the function $\partial_1 C_2^{Gau}(u_1, u_2; \rho)$ with respect to the second argument $u_2$.
It is given by
\begin{align*}
\big[\partial_1 C_2^{Gau}\big]_{2}^{-1}(u_1, u_2; \rho)
= \Phi\left( \Phi^{-1}(u_2) \sqrt{1 - \rho} + \rho \Phi^{-1}(u_1) \right).
\end{align*}
\autoref{algo_sim_dvine} can now be used to obtain a sample $(U_{1:d}, \widetilde{U}_{1,d}) \sim C_{d}^{Gau}(R)$ based on independent standard uniform random variables $(W_{1:d}, \widetilde{W}_{1,d}) \sim \mathcal{U}([0, 1]^{2d})$.

Usually, neither the marginal distributions $F_1, \ldots, F_d$ nor the parameter, or correlation, matrix $R$ are known but instead need to be estimated from data.
\autoref{algo_ko_gau} summarizes the steps to obtain Gaussian copula knockoffs.
\begin{algorithm}
\caption{Gaussian copula knockoffs}
\label{algo_ko_gau}
\DontPrintSemicolon
\KwData{$X_{1:d}$ with $X_1 \sim F_1, \ldots, X_d \sim F_d$ and $U_{1:d} := \big(F_1(X_1), \ldots, F_d(X_d)\big) \sim C_{d}^{Gau}(R)$}
\KwResult{Gaussian copula knockoffs $\widetilde{X}_{1:d}$}
\textbf{1.} Estimate marginal distributions $F_1, \ldots, F_d$\;
\textbf{2.} Compute $U_{1:d} := \big(F_1(X_1), \ldots, F_d(X_d)\big)$\;
\textbf{3.} Estimate copula parameters $R$\;
\textbf{4.} Compute partial correlations $P_{R} \longleftarrow \texttt{pcorr}(R)$\;
\textbf{5.} Compute PITs $W_{1:d} := \big( U_{1}, U_{2|1}, U_{3|1:2}, \ldots, U_{d|1:d-1} \big)$ (\autoref{algo_pits_dvine}) with $C_{d}^{Gau}(R)$\;
\textbf{6.} Find $r$ via optimization and compute $H$ (see \eqref{eq_gau_cop_ko})\;
\textbf{7.} Compute partial correlations $P_{H} \longleftarrow \texttt{pcorr}(H)$\;
\textbf{8.} Sample $\widetilde{W}_{1:d} \sim \mathcal{U}([0, 1]^d)$\;
\textbf{9.} Apply inverse PITs (\autoref{algo_sim_dvine}) to $(W_{1:d}, \widetilde{W}_{1:d})$ with $C_{2d}^{Gau}(H)$\;
\textbf{10.} Compute knockoffs $\widetilde{X}_{1:d} := (F_1^{-1}(\widetilde{U}_1), \ldots, F_d^{-1}(\widetilde{U}_d))$\;
\end{algorithm}

%% file: pcorr_vines.tex
\subsection{Partial correlation vines and completion problems}\label{sec_partial_vines}
A straightforward approach to generate Gaussian copula knockoffs relies on partial correlation vines and vine copulas.
Before we discuss partial correlation vines themselves, we want to introduce vines and the subclasses of so-called regular vines (R-vines) and drawable vines (D-vines).
Vines as graph-theoretic concept have been introduced by \textcite{bedford2002} and form the basis for partial correlation vines \parencite{bedford2002,kurowicka2003} and vine copulas \parencite{aas2009,bedford2001,joe1997}.

For $d$ variables, a vine $\mathcal{V} := (T_1, \ldots, T_{d-1}):= ((N_1, E_1), \ldots, (N_{d-1}, E_{d-1}))$ consists of $d-1$ trees.
Each tree $T_{j}$ consists of nodes $N_j$ and edges $E_j$ which form a connected graph with no cycle.
In the vine, nodes of tree $T_j$ are edges in tree $T_{j-1}$, i.e., $N_j = E_{j-1}$ for $j=2,\ldots,d-1$.
The nodes of the first tree are $N_1 = \{1, \ldots, d\}$, i.e., the variable indices.

If a vine satisfies the so-called proximity condition, it is called a regular vine (R-vine).
The proximity condition for $j=2, \ldots, d-1$ is the requirement that two nodes can only be connected by an edge in tree $T_j$, if the nodes (being edges in $T_{j-1}$) share a common node in tree $T_{j-1}$.

\begin{figure}[ht]
\begin{center}
\includegraphics[width=0.7\textwidth]{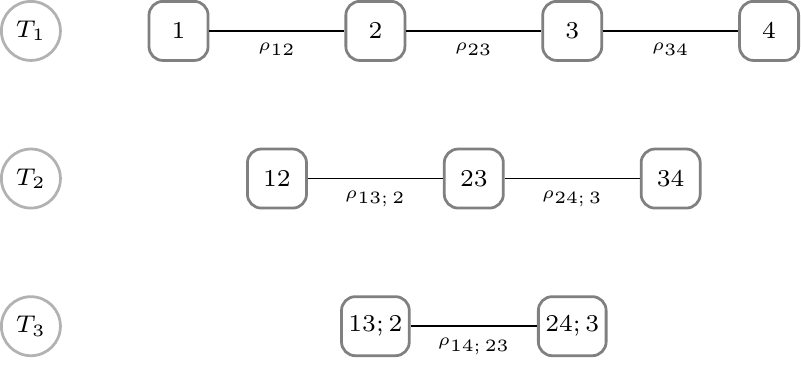}
\end{center}
\caption{Four-dimensional partial correlation D-vine.}
\label{fig_dvine}%
\end{figure}

An R-vine is called a drawable vine (D-vine), if all nodes in the first tree $T_1$ are connected to a maximum of two other nodes.
The graph-theoretic structure of a D-vine can be nicely visualized.
In \autoref{fig_dvine} we show the five-dimensional case.
The order of the variables in the first tree $T_1$ ($X_1\rightarrow \ldots\rightarrow X_d$) can be chosen arbitrarily and is sometimes also called the structure of the D-vine.

To understand the labeling of the nodes and edges in \autoref{fig_dvine}, we need to introduce some more graph-theoretic concepts and notation.
We consider an R-vine for $d$ variables.
The complete union of an edge $e = \{a, b\} \in E_j$ is defined as
\begin{align*}
\mathcal{U}_{e} := \left\lbrace n \in N_1: \exists e_1 \in E_1, \ldots, \exists e_{j-1} \in E_{j-1},\text{ with } n \in e_1 \in \ldots \in e_{j-1} \in e \right\rbrace
\end{align*}
and the conditioning set $D_e$ of the edge $e = \lbrace a, b \rbrace \in E_{j}$ in tree $T_j$ is given by $D_e := \mathcal{U}_{a} \cap \mathcal{U}_{b}$.
The conditioned sets of edge $e = \lbrace a, b \rbrace \in E_{j}$ in tree $T_j$ are defined as $h_e := \mathcal{U}_{a} \setminus D_e$ and $i_e := \mathcal{U}_{b} \setminus D_e$ and are by construction singleton indices.
The constraint set of the R-vine is defined as
$\mathcal{CV} := \lbrace (h_e, i_e\ps D_e): e \in E_j, 1 \leq j \leq d-1 \rbrace$
and contains all edge labels of the vine.\footnote{
To be more precise, the constraint set contains the indices of the partial correlations (copulas) which are assigned to the edges of the partial correlation vine (vine copula).}
For $d$ variables, the constraint set of a D-vine is given by
$\mathcal{CV} := \lbrace (j, i\ps S_{j,i}): 2\leq i \leq d$, $1 \leq j \leq i-1 \rbrace$,
where we use the short notation $S_{j,i}:= j+1:i-1 := \lbrace j+1, \ldots, i-1 \rbrace$ for the conditioning set
and the notational convention that $S_{i-1,i} := i:i-1 := \emptyset$.

In order to introduce partial correlation vines, we first want to recap the concept of partial correlations.
Consider two random variables $Y$ and $Z$ together with the vector of random variables $X$.
Further, denote by $\mathcal{P}(Y|X)$, $\mathcal{P}(Z|X)$ the unbiased projections of $Y$ on $X$ and $Z$ on $X$, respectively.
The unbiased projections can be obtained via OLS regressions.
The residuals after subtracting the projections are denoted by $\mathcal{E}_{Y|X}$, $\mathcal{E}_{Z|X}$ and given by
\begin{align*}
\mathcal{E}_{Y|X} := Y - \mathcal{P}(Y|X), \qquad\text{and}\qquad
\mathcal{E}_{Z|X} := Z - \mathcal{P}(Z|X).
\end{align*}
The partial correlation between $Y$ and $Z$ given $X$ is then defined as
\begin{align*}
\rho_{YZ\ps X} = \text{Corr}(\mathcal{E}_{Y|X}, \mathcal{E}_{Z|X})
= \frac{\mathbb{E}(\mathcal{E}_{Y|X} \mathcal{E}_{Z|X})}{\sqrt{\mathbb{E}(\mathcal{E}_{Y|X}^2) \mathbb{E}(\mathcal{E}_{Z|X}^2)}}.
\end{align*}

So-called \textit{partial correlation vines} are a specific way to parametrize multivariate Gaussian distributions \parencite{bedford2002,kurowicka2003}.
A partial correlation vine is obtained by assigning the partial correlation of $Y_{h_e}$ and $Y_{i_e}$ given $Y_{D_e}$ (with $Y_i = \Phi^{-1}(U_i)$, $i=1,\ldots, d$) to the edges of an R-vine.
If we again consider the special case of the D-vine, we can display all partial correlations assigned to the edges of the D-vine (see \autoref{fig_dvine}) in terms of a symmetric partial correlation matrix
\begin{align*}
P = 
\left(\begin{matrix}
1  & \rho_{1,2} & \rho_{1,3\ps 2} & \multicolumn{2}{c}{\cdots} & \rho_{1,d\ps 2:d-1} \\
 & 1 & \rho_{2,3} & \rho_{2,4\ps 3} & \cdots & \rho_{2,d\ps 3:d-1} \\
 & & 1 & \rho_{3,4} & \cdots & \rho_{3,d\ps 4:d-1} \\[7pt]
 & &  & \ddots & \ddots  & \vdots \\[7pt]
 & & & & 1 & \rho_{d-1,d} \\
 & & & & & 1
\end{matrix}\right),
\end{align*}
where, e.g., $\rho_{1,d\ps 2:d-1}$ denotes the partial correlation of $Y_1$ and $Y_d$ given $Y_{2:d-1}$.

A very useful property of regular vines and the corresponding partial correlation vines is that there always exists a bijective mapping between the correlation matrix $R$ and the partial correlation matrix $P$ \parencite{bedford2002}.
Therefore, the entries of the partial correlation matrix $P$ can be obtained with the bijective map from the correlation matrix $R$.
The partial correlation $\rho_{j, i\ps S_{j,i}}$ for $2\leq i \leq d$, $1 \leq j \leq i-1$, can be computed from the correlation matrix $R$ via
\begin{align*}
\rho_{j, i\ps S_{j,i}} = - \frac{Q_{1,-1}}{\sqrt{Q_{1,1} Q_{-1,-1}}},
\end{align*}
where $Q_{1,1}$ is the entry in the first row and first column of the precision matrix $Q:=R_{j:i,j:i}^{-1}=\text{Cov}(Y_{j:i})^{-1}$.
Similarly $Q_{-1,-1}$ is the entry in the last row and last column of the precision matrix $Q$ and $Q_{1,-1}$ is the entry in first row and last column.
Note that $\rho_{j, i\ps S_{j,i}}$ for $2\leq i \leq d$, $1 \leq j \leq i-1$ is the entry in the $j$-th row and $i$-th column in the upper triangular of the symmetric partial correlation matrix $P$.

Another interesting property of partial correlation vines is that the determinant of the correlation matrix $R$ can be computed from the partial correlations via \parencite{kurowicka2006}
\begin{align*}
\text{det}(R) = \prod_{j=1}^{d-1} \prod_{e \in E_j} (1 - \rho_{h_e, i_e\ps D_e}^2).
\end{align*}
This especially means that the partial correlations $\rho_{h_e, i_e\ps D_e}$, $e \in E_j$, $j=1, \ldots, d-1$, can be chosen arbitrarily in the interval $(-1, 1)$ to obtain a positive definite correlation matrix $R$ via the bijective mapping.

Partial correlation vines are well-suited to solve completion problems \parencite{kurowicka2003,kurowicka2006}.
A completion problem consists of an incompletely specified correlation matrix $R$ where the missing entries should be found under the constraint of positive definiteness.
Finding the vector $r$ for Gaussian (copula) knockoffs (see \autoref{sec_gaucop_gaucop_ko}) is exactly such a completion problem where the $d$-th off-diagonal, i.e., the correlations between $X_i$ and $\widetilde{X}_i$, is unspecified while all other correlations for the matrix $H$ are specified (see \eqref{eq_gau_cop_ko}).

We choose to work with a specific D-vine for Gaussian (copula) knockoffs because this vine structure maximizes the number of partial correlations that can be directly determined from the pairwise correlation matrix $R$ without having to choose $r$.
We highlight this fact using the four-dimensional case.
The goal is to complete the symmetric matrix $H$ given by
\begin{align*}
H =
\left(\begin{matrix}
1 & \rho_{12} & \rho_{13} & \rho_{14} &
\boldsymbol{\rho_{1\widetilde{1}}} & \rho_{1\widetilde{2}} & \rho_{1\widetilde{3}} & \rho_{1\widetilde{4}} \\
 & 1 & \rho_{23} & \rho_{24} &
\rho_{2\widetilde{1}} & \boldsymbol{\rho_{2\widetilde{2}}} & \rho_{2\widetilde{3}} & \rho_{2\widetilde{4}} \\
 & & 1 & \rho_{34} &
\rho_{3\widetilde{1}} & \rho_{3\widetilde{2}} & \boldsymbol{\rho_{3\widetilde{3}}} & \rho_{3\widetilde{4}} \\
 & & & 1 &
\rho_{4\widetilde{1}} & \rho_{4\widetilde{2}} & \rho_{4\widetilde{3}} & \boldsymbol{\rho_{4\widetilde{4}}} \\
 & & & &
1 & \rho_{\widetilde{1}\widetilde{2}} & \rho_{\widetilde{1}\widetilde{3}} & \rho_{\widetilde{1}\widetilde{4}} \\
 & & & &
 & 1 & \rho_{\widetilde{2}\widetilde{3}} & \rho_{\widetilde{2}\widetilde{4}} \\
 & & & &
 & & 1 & \rho_{\widetilde{3}\widetilde{4}} \\
 & & & &
 & & & 1 \\
\end{matrix}\right)
=
\left(\begin{matrix}
1 & \rho_{12} & \rho_{13} & \rho_{14} & \square & \rho_{12} & \rho_{13} & \rho_{14} \\
 & 1 & \rho_{23} & \rho_{24} & \rho_{12} & \square & \rho_{23} & \rho_{24} \\
 & & 1 & \rho_{34} & \rho_{13} & \rho_{23} & \square & \rho_{34} \\
 & & & 1 & \rho_{14} & \rho_{24} & \rho_{34} & \square \\
 & & & & 1 & \rho_{12} & \rho_{13} & \rho_{14} \\
 & & & & & 1 & \rho_{23} & \rho_{24} \\
 & & & & & & 1 & \rho_{34} \\
 & & & & & & & 1 \\
\end{matrix}\right).
\end{align*}
Given the symmetric pairwise-correlation matrix $H$, we can directly compute the partial correlations on the first $d-1$ off-diagonals, i.e., in the four-dimensional case the first three off-diagonals of the partial correlation matrix
\begin{align*}
P &= 
\left(\begin{matrix}
1 & \rho_{12} & \rho_{13\ps 2} & \rho_{14\ps 23} &
\boldsymbol{\rho_{1\widetilde{1}\ps 234}} & \rho_{1\widetilde{2}\ps 234\widetilde{1}} & \rho_{1\widetilde{3}\ps 234\widetilde{1}\widetilde{2}} & \rho_{1\widetilde{4}\ps 234\widetilde{1}\widetilde{2}\widetilde{3}} \\
 & 1 & \rho_{23} & \rho_{24\ps 3} &
\rho_{2\widetilde{1}\ps 34} & \boldsymbol{\rho_{2\widetilde{2}\ps 34\widetilde{1}}} & \rho_{2\widetilde{3}\ps 34\widetilde{1}\widetilde{2}} & \rho_{2\widetilde{4}\ps 34\widetilde{1}\widetilde{2}\widetilde{3}} \\
 & & 1 & \rho_{34} &
\rho_{3\widetilde{1}\ps 4} & \rho_{3\widetilde{2}\ps 4\widetilde{1}} & \boldsymbol{\rho_{3\widetilde{3}\ps 4\widetilde{1}\widetilde{2}}} & \rho_{3\widetilde{4}\ps 4\widetilde{1}\widetilde{2}\widetilde{3}} \\
 & & & 1 &
\rho_{4\widetilde{1}} & \rho_{4\widetilde{2}\ps \widetilde{1}} & \rho_{4\widetilde{3}\ps \widetilde{1}\widetilde{2}} & \boldsymbol{\rho_{4\widetilde{4}\ps \widetilde{1}\widetilde{2}\widetilde{3}}} \\
 & & & &
1 & \rho_{\widetilde{1}\widetilde{2}} & \rho_{\widetilde{1}\widetilde{3}\ps \widetilde{2}} & \rho_{\widetilde{1}\widetilde{4}\ps \widetilde{2}\widetilde{3}} \\
 & & & &
 & 1 & \rho_{\widetilde{2}\widetilde{3}} & \rho_{\widetilde{2}\widetilde{4}\ps \widetilde{3}} \\
 & & & &
 & & 1 & \rho_{\widetilde{3}\widetilde{4}} \\
 & & & &
 & & & 1 \\
\end{matrix}\right) \\
&=
\left(\begin{matrix}
1 & \rho_{12} & \rho_{13\ps 2} & \rho_{14\ps 23} & \square & \square & \square & \square \\
 & 1 & \rho_{23} & \rho_{24\ps 3} & \rho_{21\ps 34} & \square & \square & \square \\
 & & 1 & \rho_{34} & \rho_{31\ps 4} & \rho_{32\ps 41} & \square & \square \\
 & & & 1 & \rho_{41} & \rho_{42\ps 3} & \rho_{43} & \square \\
 & & & & 1 & \rho_{12} & \rho_{13\ps 2} & \rho_{14\ps 23} \\
 & & & & & 1 & \rho_{23} & \rho_{24\ps 3} \\
 & & & & & & 1 & \rho_{34} \\
 & & & & & & & 1 \\
\end{matrix}\right).
\end{align*}

It is an interesting idea for future research to solve such completion problems with partial correlation vines in order to determine the matrix $H$ for Gaussian (copula) knockoffs.
However, one can also use the standard optimization approach to obtain $r$ depending on $R$ (see \autoref{sec_gaucop_gaucop_ko}).
Having specified $r$ and therefore $H$, all partial correlations of the D-vine can be computed via the bijective mapping.
In this work we instead want to put the focus on generalizations to more flexible dependence structures.
These are obtained with vine copulas which generalize partial correlation vines by assigning copulas to the edges of R-vines instead of partial correlations.

%% file: vine_cop.tex
\subsection{Vine copulas and the simplifying assumption}
Vine copulas allow to represent multivariate copulas in terms of bivariate copulas that are assigned to the edges of an R-vine and have been introduced by \textcite{aas2009,joe1997,bedford2002}.
To represent an arbitrary $d$-dimensional copula $C_{1:d}$ in terms of an R-vine copula, we need to introduce conditional probability integral transforms and conditional copulas.
For $U_{1:d}$, the (conditional) probability integral transform (PIT) of $U_{l_e}$ given $U_{D_e}$ is defined as $U_{l_e|D_e}:= F_{l_e|D_e}(U_{l_e}|U_{D_e})$, $l_e = h_e, i_e$, $e\in E_j$, $j=2, \ldots, d-1$.
PITs are by construction uniformly distributed.
The conditional copula \parencite{patton2006} is defined as the conditional joint distribution of the PITs $(U_{h_e|D_e}, U_{i_e|D_e})$ given $U_{D_e}$ and denoted by $C_{h_e, i_e\ps D_e}$.
The R-vine copula representation \parencite{bedford2001} of a multivariate copula $C_{1:d}$ is now obtained by assigning the unconditional bivariate copulas $C_{h_e, i_e}$ to the edges in the first tree $e \in E_1$ and the conditional bivariate copulas $C_{h_e, i_e\ps D_e}$ to the edges $e \in E_j$, $j=2, \ldots, d-1$ of higher trees.
It can be shown that any copula density $c_{1:d}$ can then be factorized as
\begin{align*}
c_{1:d}(u_{1:d}) = \prod_{j=1}^{d-1} \prod_{e \in E_j} c_{h_e, i_e\ps D_e}(u_{h_e|D_e}, u_{i_e|D_e}| u_{D_e}),
\end{align*}
where $u_{l_e|D_e} = F_{l_e|D_e}(u_{l_e}|u_{D_e})$, $l_e = h_e, i_e$, $e\in E_j$, $j=2, \ldots, d-1$.
For simplicity, in the first tree we further set $u_{l_e|D_e} = u_{l_e}$ for $l_e = h_e, i_e$, $e\in E_1$
and $C_{h_e, i_e\ps D_e} = C_{h_e, i_e}$ for $e\in E_1$.
We collect all $d(d-1)/2$ bivariate copulas determining the vine copula model in the set
\begin{align*}
\mathcal{C}_{d}^{\mathcal{V}}
&= \big\lbrace C_{h_e, i_e\ps D_e}: e \in E_j, j=1, \ldots, d-1 \big\rbrace.
\end{align*}

The estimation of such a $d$-dimensional vine copula is challenging in high-dimensional settings.
To overcome the curse of dimensionality in the modeling process, one usually assumes that the conditional copulas $c_{h_e, i_e\ps D_e}(\cdot, \cdot| u_{D_e})$ do not vary in $u_{D_e}$ for $e\in E_j$, $j=2, \ldots, d-1$.
This modeling assumption is called the simplifying assumption \parencite{hobakhaff2010}.
For discussions, analysis and a statistical test of the simplifying assumption see \textcite{hobakhaff2010,stoeber2013,mroz2021,spanhel2019,kurz2021}.

The simplifying assumption is satisfied for the multivariate Gaussian copula \parencite{hobakhaff2010,stoeber2013}.
As a consequence, the density of a $d$-dimensional Gaussian copula $C_{d}^{Gau}(R)$ can be written as the product of $d(d-1)/2$ bivariate Gaussian copula densities
\begin{align*}
c_{1:d}^{Gau}(u_{1:d}; R) = \prod_{j=1}^{d-1} \prod_{e \in E_j} c_{h_e, i_e\ps D_e}^{Gau}(u_{h_e|D_e}, u_{i_e|D_e}\ps \rho_{h_e, i_e\ps D_e}),
\end{align*}
where the parameters of the bivariate Gaussian copulas are the partial correlations from the corresponding partial correlation vine, i.e.,
$\mathcal{C}_{d}^{\mathcal{V}}
= \big\lbrace C_{h_e, i_e\ps D_e}^{Gau}(\cdot, \cdot\ps \rho_{h_e, i_e\ps D_e}) : e \in E_j, j=1, \ldots, {d-1} \big\rbrace$.

A key advantage of vine copulas is that there exist straightforward algorithms to compute things like probability integral transforms or to simulate from vine copulas using an iterative approach.
These computations only involve evaluations of bivariate copula functions and derivatives thereof, which renders the algorithms computationally feasible.

In the following, we show \autoref{algo_pits_dvine} which can be used to compute the probability integral transforms
\begin{align*}
W_{1:d} &:= \big( U_1, U_{2|1}, U_{3|1:2}, \ldots, U_{d|1:d-1}\big) \\
&:= \big( F_1(U_1), F_{2|1}(U_2|U_1), F_{3|1:2}(U_3|U_{1:2}), \ldots, F_{d|1:d-1}(U_d|U_{1:d-1}) \big)
\end{align*}
from a simplified D-vine copula $\mathcal{C}_{d}^{\mathcal{V}}$.
To apply \autoref{algo_pits_dvine}, all bivariate copulas assigned to the edges of the D-vine need to be specified.
In addition, conditional distribution functions for these copulas are required.
They can be obtained via partial derivatives.
If $(U_{1}, U_{2}) \sim C$, the conditional distributions are given by
\begin{align*}
F_{U_1|U_2}(u_1| u_2) &= \partial_2 C(u_1, u_2), \\
F_{U_2|U_1}(u_2| u_1) &= \partial_1 C(u_1, u_2),
\end{align*}
where we use the notation $\partial_i f(x)$ for the partial derivative of a function $f(x)$ with respect to the $i$-th variable $x_i$.
\begin{algorithm}
\caption{Compute PITs from a simplified D-vine copula}
\label{algo_pits_dvine}
\DontPrintSemicolon
\KwData{$U_{1:d} \sim C_{1:d}$ and a simplified D-vine copula $\mathcal{C}_{d}^{\mathcal{V}}$}
\KwResult{$W_{1:d} := \big( U_{1}, U_{2|1}, U_{3|1:2}, \ldots, U_{d|1:d-1} \big)$}
$W_1, a_1, b_1 \longleftarrow U_1$\;
\For{$i\leftarrow 2$ \KwTo $d$}{
$a_i \longleftarrow U_i$\;
\For{$j\leftarrow i-1$ \KwTo $1$}{
$a_{j} \longleftarrow \partial_1 C_{j, i\ps S_{j,i}}(b_j, a_{j+1}; \theta_{j,i})$\;
}
$W_i \longleftarrow a_1$;
$b_i \longleftarrow a_i$\;
\For{$j\leftarrow 1$ \KwTo $i-1$}{
$b_{j} \longleftarrow \partial_2 C_{j, i\ps S_{j,i}}(b_j, a_{j+1}; \theta_{j,i})$\;
}
}
\end{algorithm}

A random sample from a copula or vine copula can be obtained by using inverse probability integral transforms.
This method is also often called inverse probability method for random number generation.
For this, we define the inverse of the conditional distribution function $F_{U_2|U_1}(u_2| u_1)$ with respect to $u_2$.
It is denoted by
\begin{align*}
F_{U_2|U_1}^{-1}(u_2| u_1) = \big[\partial_1 C\big]_{2}^{-1}(u_1, u_2),
\end{align*}
where we use the notation $f_i^{-1}$ for the inverse of a function $f$ with respect to the $i$-th variable.
\autoref{algo_sim_dvine} can be used to obtain a sample $U_{1:d}$ from a $d$-dimensional D-vine copula based on independent standard uniform random variables $W_{1:d} \sim \mathcal{U}([0, 1]^d)$.
\begin{algorithm}
\caption{Simulation from a simplified D-vine copula}
\label{algo_sim_dvine}
\DontPrintSemicolon
\KwData{$W_{1:d} \sim \mathcal{U}([0, 1]^d)$ and a simplified D-vine copula $\mathcal{C}_{d}^{\mathcal{V}}$}
\KwResult{$U_{1:d} \sim C_{1:d}$}
$U_1, a_1, b_1 \longleftarrow W_1$\;
\For{$i\leftarrow 2$ \KwTo $d$}{
$a_1 \longleftarrow W_i$\;
\For{$j\leftarrow 1$ \KwTo $i-1$}{
$a_{j+1} \longleftarrow \big[\partial_1 C_{j, i\ps S_{j,i}}\big]_{2}^{-1}(b_j, a_j; \theta_{j,i})$\;
}
$U_i, b_i \longleftarrow a_i$\;
\For{$j\leftarrow 1$ \KwTo $i-1$}{
$b_{j} \longleftarrow \partial_2 C_{j, i\ps S_{j,i}}(b_j, a_{j+1}; \theta_{j,i})$\;
}
}
\end{algorithm}

%% file: vine_ko.tex
\section{Vine copula knockoffs: Generating knockoffs with D-vine copulas}\label{sec_vine_ko}
Gaussian copula knockoffs are more flexible than Gaussian knockoffs because the model class allows for arbitrary marginal distributions $F_1, \ldots, F_d$ and is not restricted to normally distributed margins.
However, the dependence structure of $(X_{1:d}, \widetilde{X}_{1:d})$ is restricted to a $2d$-dimensional Gaussian copula.
We have already seen that a Gaussian copula can be parametrized in terms of a vine copula by assigning bivariate Gaussian copulas to the edges of the vine with parameters given by the corresponding partial correlation vine.
A straightforward generalization, allowing for much more flexible dependence structures, can be obtained by assigning more flexible copulas to the edges of the vine copula model.
The procedure to obtain knockoffs from a flexible vine copula model will be discussed in the following.
We call it \textit{vine copula knockoffs} and an implementation thereof is provided in the \texttt{Python} package \texttt{vineknockoffs} \parencite{kurz2022}.

As for the Gaussian copula knockoffs, the marginal distributions $F_1, \ldots, F_d$ can be estimated separately from the dependence structure.
Given the marginal distributions, we transform $X_{1:d}$ to standard uniformly distributed random variables $U_{1:d} := \big(F_1(X_1), \ldots, F_d(X_d)\big)$.
The first step for estimating a vine copula model is the selection of the vine structure.
We choose the previously discussed D-vine because it maximizes the number of bivariate copulas that can be directly estimated from the data.
The order of the variables in a D-vine can be chosen arbitrarily.
The most common heuristic procedure suggested in the vine copula literature to determine the order is given by maximizing the dependence in the first tree of the D-vine \parencite[see for example][]{dissmann2013}.
This is obtained by finding the shortest Hamiltonian path using $1-|\tau_{ij}|$ as weights, where $\tau_{ij}$ is the pairwise Kendall's $\tau$ of variables $X_i$ and $X_j$.
The optimization problem to find the order can be solved via the standard approaches for traveling salesman problems (TSP).
We apply this procedure to determine the order of the variables $X_{1:d}$ and use the same order for the knockoffs $\widetilde{X}_{1:d}$.
For notational convenience, we assume in the following that the order obtained by solving the TSP is
$X_1\rightarrow X_2\rightarrow \ldots \rightarrow X_{d-1}\rightarrow X_d$.

\begin{figure}[ht]
\includegraphics[width=0.95\textwidth]{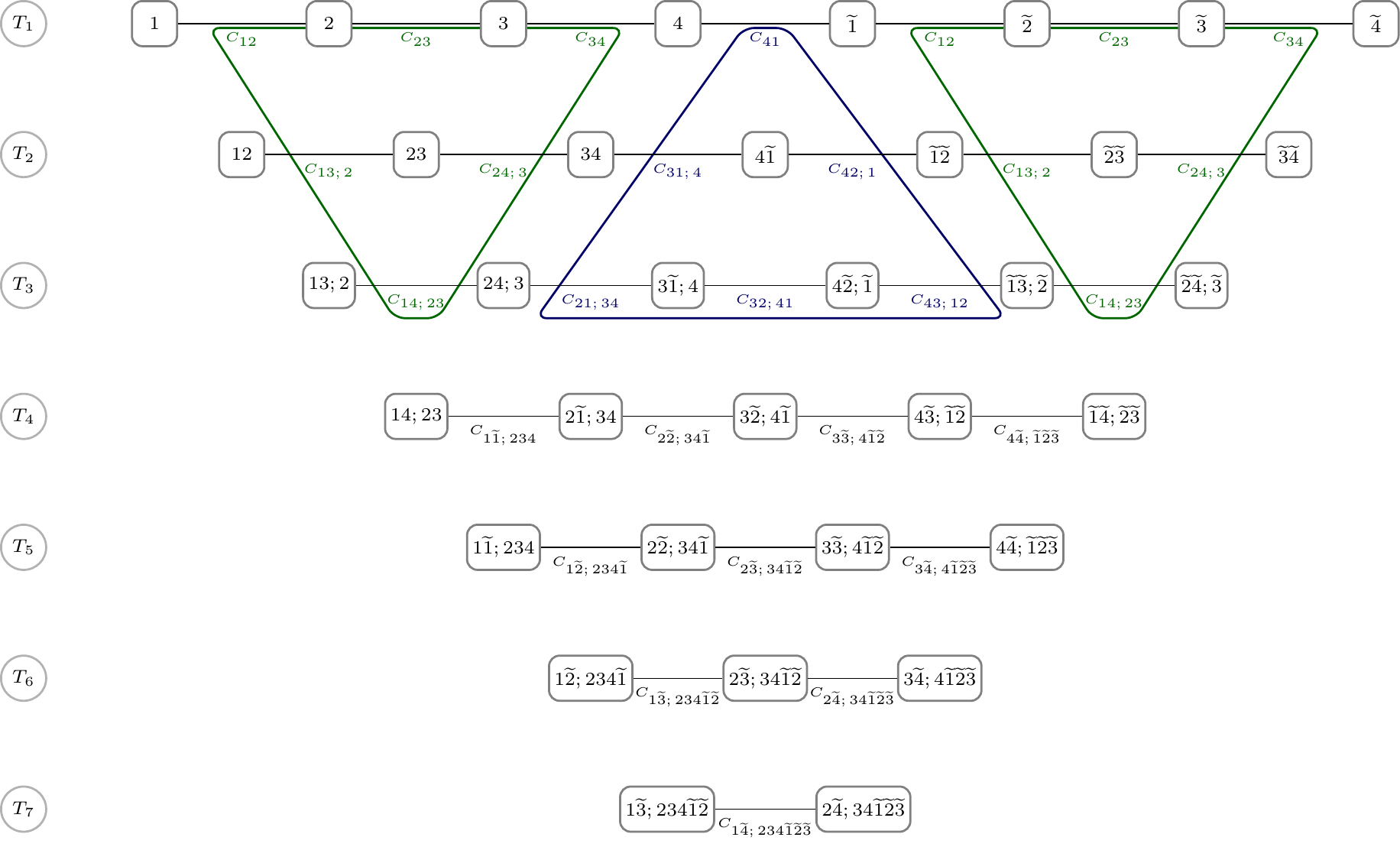}
\caption{Vine copula knockoffs: D-vine copula for knockoff generation with $d=4$.}
\label{fig_dvine_ko}
\end{figure}

The next step in vine copula modeling is the specification of bivariate copulas for each edge of the D-vine copula of $(U_{1:d}, \widetilde{U}_{1:d})$.
Following the same arguments as for the partial correlation D-vine, we can estimate the bivariate copulas in trees $T_1$ to $T_{d-1}$ directly from the data.
In the following, we work with parametric copula families (Gaussian, Clayton, Frank, Gumbel, Independence) and select the families with the Akaike information criterion (AIC) while estimating the parameters with maximum likelihood.
Due to the special D-vine structure, there are $d(d-1)/2$ copulas that can each be assigned to two edges.
These are highlighted in green in \autoref{fig_dvine_ko}, which shows the four-dimensional case, i.e., a eight-dimensional D-vine copula as model for the distribution of $(U_{1:4}, \widetilde{U}_{1:4})$.
By mirroring the copulas which determine the joint distribution of $U_{1:d}$, we achieve that the model for the joint distribution of $\widetilde{U}_{1:d}$ is exactly the same as the model for the joint distribution of $U_{1:d}$.
Furthermore, it is computationally more efficient to estimate the copulas only once and then assign them to two edges.
A second set of bivariate copulas can be estimated directly from the data.
These $d(d-1)/2$ copulas are highlighted in blue in \autoref{fig_dvine_ko}.
From a computational point of view, one can simply duplicate the data to obtain the $2d$-dimensional vector $(U_{1:d}, U_{1:d})$ and then estimate the copulas in tree $T_1$ to $T_{d-1}$ with the standard algorithm for a $2d$-dimensional D-vine copula \parencite[see for example Algorithm 3.2 in][]{cooke2010}.
However, in each tree $T_1, \ldots, T_{d-1}$ only the first $d$ copulas (from left to right in \autoref{fig_dvine_ko}) need to be estimated and the remaining copulas are obtained via the above described mirroring approach.

Next the copulas in tree $T_d$ need to be specified.
In the four-dimensional case, visualized in \autoref{fig_dvine_ko}, this are the copulas $C_{1\widetilde{1}\ps 234}$, $C_{2\widetilde{2}\ps 34\widetilde{1}}$, $C_{3\widetilde{3}\ps 4\widetilde{1}\widetilde{2}}$ and $C_{4\widetilde{4}\ps \widetilde{1}\widetilde{2}\widetilde{3}}$.
These copulas fundamentally determine the dependence between each variable $X_{i}$ and its knockoff $\widetilde{X}_i$.
The joint density of $(X_{1:4}, \widetilde{X}_1)$ is for example given by (assuming the simplifying assumption holds)
\begin{align*}
f_{1234\widetilde{1}}(x_{1:4}, \widetilde{x}_1) = c_{1\widetilde{1}\ps 234}(F_{1|2:4}(x_1| x_{2:4}), F_{\widetilde{1}|2:4}(x_{\widetilde{1}}| x_{2:4}))
\cdot f_{1|2:4}(x_1| x_{2:4}) \cdot f_{\widetilde{1}|2:4}(x_{\widetilde{1}}| x_{2:4}) \cdot f_{2:4}(x_{2:4}).
\end{align*}
Obviously, the joint density of $X_{i}$ and its knockoff $\widetilde{X}_i$ could be extracted via integration from $f_{1234\widetilde{1}}$.
While it is possible to estimate the copulas in tree $T_1, \ldots, T_{d-1}$ of the D-vine directly from data, this is not recommended for higher trees.
If one would directly estimate the copulas in the higher trees from the duplicate data vector $(U_{1:d}, U_{1:d})$, one would observe the very same phenomena as for Gaussian knockoffs.
Namely, the constructed knockoffs would tend to be perfectly positive dependent with the original variables.
However, to have power one usually tries to obtain knockoffs which are as independent as possible from the original variables.
We therefore suggest the following heuristic to specify the copulas in the trees $T_{d}, \ldots, T_{2d-1}$:
Instead of continuing the estimation of copulas in an iterative fashion, we calculate the partial correlation D-vine from matrix $H$ as described for the Gaussian copula knockoffs.
We then assign bivariate Gaussian copulas with the corresponding partial correlations $\rho_{j,i\ps S_{j,i}}$ to the edges in the trees $T_{d}, \ldots, T_{2d-1}$.\footnote{
The heuristically obtained parameters might be further optimized.
In our outlook and \autoref{sec_sgd}, we discuss a stochastic gradient descent algorithm that could be used to further improve and tune vine copula knockoff models.}

The procedure to generate vine copula knockoffs can be summarized as follows:
First we transform $X_{1:d}$ to standard uniformly distributed random variables $U_{1:d} := \big(F_1(X_1), \ldots, F_d(X_d)\big)$.
Then we determine the order of the variables, which specifies the structure of the D-vine.
Next, the copulas in the trees $T_1, \ldots, T_{d-1}$ are estimated from the data with the standard algorithm for a $2d$-dimensional D-vine copula model.
However, we can exploit the fact that $d(d-1)/2$ copulas can be assigned to two edges each.
We further compute the matrix $H$ (see \eqref{eq_gau_cop_ko}) and the corresponding partial correlation vine.
In the trees $T_{d}, \ldots, T_{2d-1}$, we assign Gaussian copulas with the corresponding partial correlations $\rho_{j,i\ps S_{j,i}}$ as parameters to the edges of the D-vine copula.
We obtain the D-vine copula model $\mathcal{C}_{2d}^{\mathcal{V}}$.
Then we compute the PITs $W_{1:d} := \big( U_{1}, U_{2|1}, U_{3|1:2}, \ldots, U_{d|1:d-1} \big)$ from $U_{1:d}$ using \autoref{algo_pits_dvine}.
Next, we sample $\widetilde{W}_{1:d} \sim \mathcal{U}([0, 1]^d)$ and apply the inverse probability integral transform (\autoref{algo_sim_dvine}) to $(W_{1:d}, \widetilde{W}_{1:d})$ with the $2d$-dimensional vine copula model $\mathcal{C}_{2d}^{\mathcal{V}}$.
As a result we obtain $(U_{1:d}, \widetilde{U}_{1,d})$ from the conditional distribution $F_{\widetilde{U}_{1:d}|U_{1:d}}$.
Finally, we set $\widetilde{X}_{1:d} := (F_1^{-1}(\widetilde{U}_1), \ldots, F_d^{-1}(\widetilde{U}_d))$
in order to obtain knockoffs $\widetilde{X}_{1:d}$ as a sample from the conditional distribution $F_{\widetilde{X}_{1:d}|X_{1:d}}$.
\autoref{algo_ko_vine} summarizes the steps to obtain vine copula knockoffs.
\begin{algorithm}
\caption{Vine copula knockoffs}
\label{algo_ko_vine}
\DontPrintSemicolon
\KwData{$X_{1:d}$}
\KwResult{Vine copula knockoffs $\widetilde{X}_{1:d}$}
\textbf{1.} Estimate marginal distributions $F_1, \ldots, F_d$\;
\textbf{2.} Compute $U_{1:d} := \big(F_1(X_1), \ldots, F_d(X_d)\big)$\;
\textbf{3.} Determine the order of variables for the D-vine\;
\textbf{4.} Estimate a D-vine copula model $\mathcal{C}_{2d}^{\mathcal{V}} := \big\lbrace C_{j, i\ps S_{j,i}}: 2\leq i \leq 2d$, $1 \leq j \leq i-1 \big\rbrace$\;
\textbf{5.} Compute PITs $W_{1:d} := \big( U_{1}, U_{2|1}, U_{3|1:2}, \ldots, U_{d|1:d-1} \big)$ (\autoref{algo_pits_dvine}) \;
\textbf{6.} Sample $\widetilde{W}_{1:d} \sim \mathcal{U}([0, 1]^d)$\;
\textbf{7.} Apply inverse PITs (\autoref{algo_sim_dvine}) to $(W_{1:d}, \widetilde{W}_{1:d})$ with $\mathcal{C}_{2d}^{\mathcal{V}}$ to obtain $(U_{1:d}, \widetilde{U}_{1:d})$\;
\textbf{8.} Compute knockoffs $\widetilde{X}_{1:d} := (F_1^{-1}(\widetilde{U}_1), \ldots, F_d^{-1}(\widetilde{U}_d))$\;
\end{algorithm}

%% file: sim.tex
\section{Simulation study}\label{sec_sim}
In the following simulation study we analyze the finite sample performance of vine copula knockoffs for high-dimensional controlled variable selection.
We will compare three different knockoff methods discussed in the previous sections: Gaussian knockoffs (\autoref{sec_gau_ko}), Gaussian copula knockoffs (\autoref{algo_ko_gau} in \autoref{sec_gaucop_ko}) and vine copula knockoffs (\autoref{algo_ko_vine} in \autoref{sec_vine_ko}).\footnote{
Note that for the Gaussian knockoffs and Gaussian copula knockoffs we use the D-vine structure $X_1\rightarrow X_2\rightarrow \ldots \rightarrow X_{d-1}\rightarrow X_d$, i.e., the same order of variables as in the data generation step itself.
For vine copula knockoffs, we assume that this order is not known and the D-vine structure is chosen in a data-driven way using the previously described approach based on solving a traveling salesman problem (see \autoref{sec_vine_ko}).
This is obviously more challenging and makes the modeling process more complex than if we would assume to know the data generating structure.}
All three knockoff methods are implemented in the \texttt{Python} package \texttt{vineknockoffs} \parencite{kurz2022}.
For the two copula-based knockoff methods, we need to estimate the marginal distributions $F_1, \ldots, F_d$.
This is done with the \texttt{R} package \texttt{kde1d} \parencite{nagler2022}, which implements a univariate kernel density estimator \parencite[see also][]{geenens2014,geenens2018,nagler2018a,nagler2018b}.

In the simulation study, we consider three different data generating processes (DGPs) for the covariates $X_{1:d} \in \mathbb{R}^d$: A multivariate normal distribution, a Gaussian copula and a truncated Clayton vine copula.
We set $d=50$ and conditionally on $X_{1:d}$ simulate the response variables from a normal distribution $Y \stackrel{iid}{\sim} \mathcal{N}(X_{1:d} \beta, 1)$.
We randomly choose twenty entries of the $d$-dimensional parameter vector $\beta$ to be non-zero.
To alter the magnitude of the effects, we use the same setup as \textcite{romano2020} and set these parameters equal to $\pm \alpha/ \sqrt{n}$, where $n$ is the sample size and $\alpha \in \lbrace 3, 6, 9, 12, 15 \rbrace$ is used to vary the magnitude.
The signs of the non-zero $\beta$-entries are sampled randomly.
As \textcite{romano2020}, we draw independent samples of size $m=10000$ from the three DGPs in order to estimate the parameters of the knockoff models.

To study the finite sample performance for controlled variable selection, we proceed in the following way.
For each $\alpha \in \lbrace 3, 6, 9, 12, 15 \rbrace$, we sample 1000 datasets of size $n=200$ from each DGP.
The three different knockoff methods are then used to obtain knockoff copies $\widetilde{X}_{1:d}$.
We then perform variable selection of $Y$ given the augmented data vector $(X_{1:d}, \widetilde{X}_{1:d})$ using the standard Lasso regression framework.
The Lasso penalty parameter $\lambda$ is determined via cross-validation using the \texttt{R} package \texttt{glmnet} \parencite{friedman2010}.
Given the Lasso parameter estimates $(\hat{\beta}, \tilde{\beta})$, we then apply the knockoff filter using the statistics $W_i = |\hat{\beta}_i| - |\tilde{\beta}_i|$ for $i \in \lbrace 1, \ldots, d \rbrace$ with a nominal level $q=0.1$.
Knowing the ground truth parameter vectors $\beta$, we can determine true and false positives.
This allows us to compare the finite sample performance of different knockoff methods by comparing the false discovery rates and empirical power results among the methods.

To further evaluate and compare the quality of the three knockoff methods, we compute the following diagnostics.
The marginal distributions of each knockoff variable $\widetilde{X}_i$ should be equal to the distribution of the original variable $X_i$, $1 \leq i \leq d$.
As diagnostic for the marginal distributions, we compute the Kolmogorov-Smirnov test statistics for each pair $(X_{i}, \widetilde{X}_{i})$ and show boxplots of the average statistics (over the $d=50$ variables) for the different knockoff methods.
To evaluate whether the joint distribution of $(X_{1:d}, \widetilde{X}_{1:d})$ is invariant with respect to swaps (knockoff property \eqref{eq_ko_1}), we further compute maximum mean discrepancies (MMDs) as suggested in \textcite{romano2020}.
The maximum mean discrepancy (MMD) is a kernel-based statistic for the equality of two distributions proposed by \textcite{gretton2012}.
To evaluate the knockoff property \eqref{eq_ko_1}, we split each sample into two groups and set $Z_1=(X_{1:d}, \widetilde{X}_{1:d})$ and $Z_2=(\widetilde{X}_{1:d}, X_{1:d})$ in the two groups, respectively.
The MMD between $Z_1$ and $Z_2$ is then a measure for the sensitivity of the joint distribution $(X_{1:d}, \widetilde{X}_{1:d})$ with respect to a full swap, i.e., a swap where every of the $d$ variables is being swapped.

\subsection{DGP: Multivariate Gaussian}
\begin{figure}[!h]
\begin{center}
\includegraphics[width=0.8\textwidth]{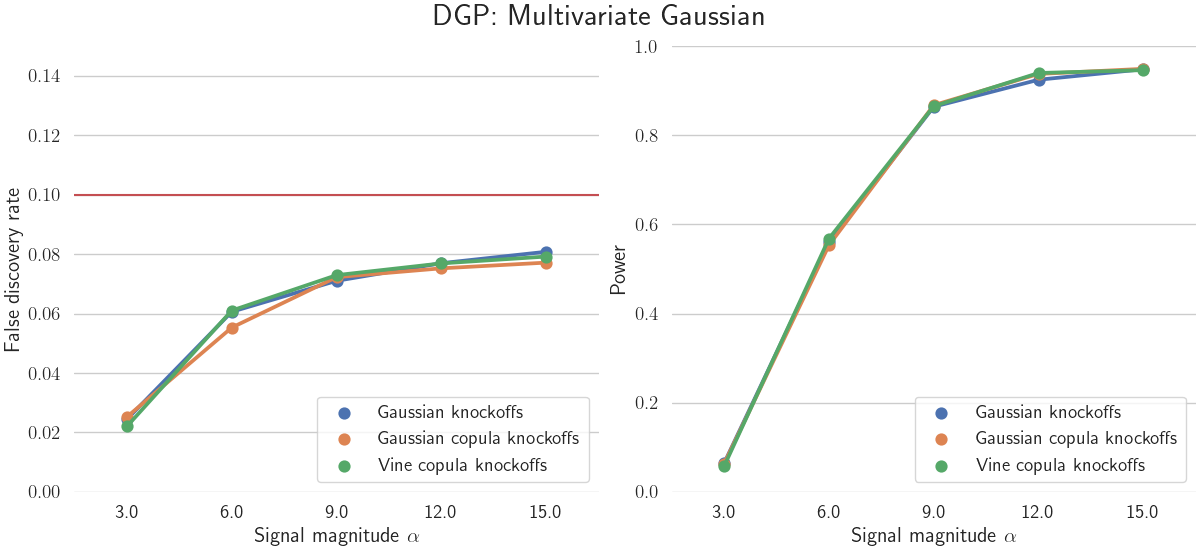}
\end{center}
\caption{
Average false discovery proportions (left) and empirical power (right) plotted against the signal magnitude $\alpha \in \lbrace 3, 6, 9, 12, 15 \rbrace$.
The DGP is the multivariate normal distribution \eqref{eq_gau_dgp} with $d=50$ and $n=200$.
The results for Gaussian knockoffs, Gaussian copula knockoffs and vine copula knockoffs are shown color-coded and obtained with $1000$ independent repetitions.
}\label{fig_ko_study_mult_gau}
\end{figure}

As first data generating process (DGP) we consider the following multivariate normal distribution
\begin{align}
X_{1:d} \sim \mathcal{N}_d(0, \Sigma), \label{eq_gau_dgp}
\end{align}
with covariance matrix $\Sigma$ being a Toeplitz matrix with entries $\Sigma_{ij} = \rho^{|i-j|}$.
We set $\rho=0.7$ and consider a sample size of $n=200$.
For each $\alpha \in \lbrace 3, 6, 9, 12, 15 \rbrace$, we repeat the experiment $1000$ times.
The results of the controlled variable selection are presented in \autoref{fig_ko_study_mult_gau}.
On the left in \autoref{fig_ko_study_mult_gau}, we plot the average false discovery proportions in $1000$ repetitions against the signal magnitude $\alpha$.
The results for the three considered knockoff methods are very similar.
Comparing the false discover proportions with the nominal level $q=0.1$, it becomes evident that with all three methods the false discovery rates can be controlled.
Also in terms of the empirical power (right hand side of \autoref{fig_ko_study_mult_gau}) all three knockoff generation methods perform similarly.
\begin{figure}[!h]
\begin{center}
\includegraphics[width=0.8\textwidth]{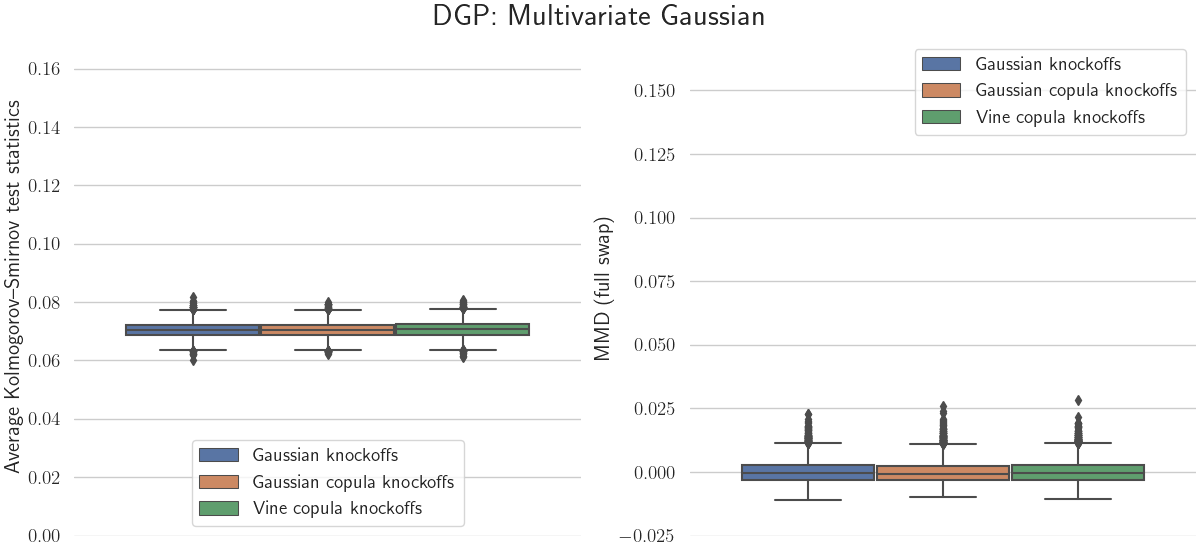}
\end{center}
\caption{
Knockoff diagnostics for the multivariate normal distribution \eqref{eq_gau_dgp} with $d=50$ and $n=200$ as DGP.
On the left boxplots of average Kolmogorov-Smirnov test statistics for the pairs $(X_i, \widetilde{X}_i)$, $1 \leq i \leq d$ are shown.
The right plot provides boxplots of the MMD statistic for a full swap.
The results for Gaussian knockoffs, Gaussian copula knockoffs and vine copula knockoffs are shown color-coded and obtained with $1000$ independent repetitions.
}\label{fig_diagnostics_mult_gau}
\end{figure}

In \autoref{fig_diagnostics_mult_gau}, we present diagnostics for the three knockoff methods and the multivariate normal distribution as DGP.
It shows that parametric normal distributions (Gaussian knockoffs) and the univariate kernel density estimators (Gaussian copula knockoffs and vine copula knockoffs) perform equally good in terms of the average Kolmogorov-Smirnov test statistics.
Looking at the maximum mean discrepancies (MMD) shown in the boxplots on the right of \autoref{fig_diagnostics_mult_gau}, we see that with all three knockoff methods the full swap property is equally well satisfied.

\subsection{DGP: Gaussian copula}\label{sec_sim_gau_cop}
The second DGP is a Gaussian copula given by
\begin{align}
U_{1:d} := \big(F_1(X_1), \ldots, F_d(X_d)\big) \sim C_{d}^{Gau}(R), \label{eq_gau_cop_dgp}
\end{align}
with correlation matrix $R$ being a Toeplitz matrix with entries $R_{ij} = \rho^{|i-j|}$.
We set $\rho=0.7$ and consider a sample size of $n=200$.
In contrast to the multivariate Gaussian DGP, we want to study more flexible marginal distribution.
Therefore, we randomly pick each margin $F_i$, $1 \leq i \leq d$, among the following distributions:
\begin{enumerate}
\item A \emph{Gaussian mixture} distribution with three equally-likely mixture components.
Each mixture component is a normal distribution $\mathcal{N}(\mu_j, \sigma_j^2)$ where the parameters are drawn randomly: $\mu_j \sim \mathcal{N}(0, 4)$ and $\sigma_j^2 \sim \mathcal{W}^{-1}(1, 3)$, $1 \leq j \leq 3$. Here, $\mathcal{W}^{-1}(\Psi, \nu)$ denotes an inverse Wishart distribution with $\nu$ degrees of freedom and scale parameter $\Psi$.
\item A \emph{Student-t} distribution with $\nu = 3$ degrees of freedom, scale parameter $\sigma$ and location parameter $\mu$, which are drawn randomly via $\mu \sim \mathcal{N}(0, 4)$ and $\sigma^2 \sim \mathcal{W}^{-1}(1, 3)$.
\item An \emph{Exponential} distribution where the parameter $\lambda$ is drawn randomly from a Gamma distribution.
\end{enumerate}
The Gaussian copula DGP is obtained by once drawing $d=50$ marginal distributions (Gaussian mixture, Student-t or Exponential).
We then sample from the Gaussian copula and obtain the random covariates $X_i = F_i^{-1}(U_i)$, $1 \leq i \leq d$.
As described above, an independent sample is used to fit the knockoff models and we then repeat the knockoff experiment with $1000$ samples of size 
$n=200$.

\begin{figure}[!htb]
\begin{center}
\includegraphics[width=0.8\textwidth]{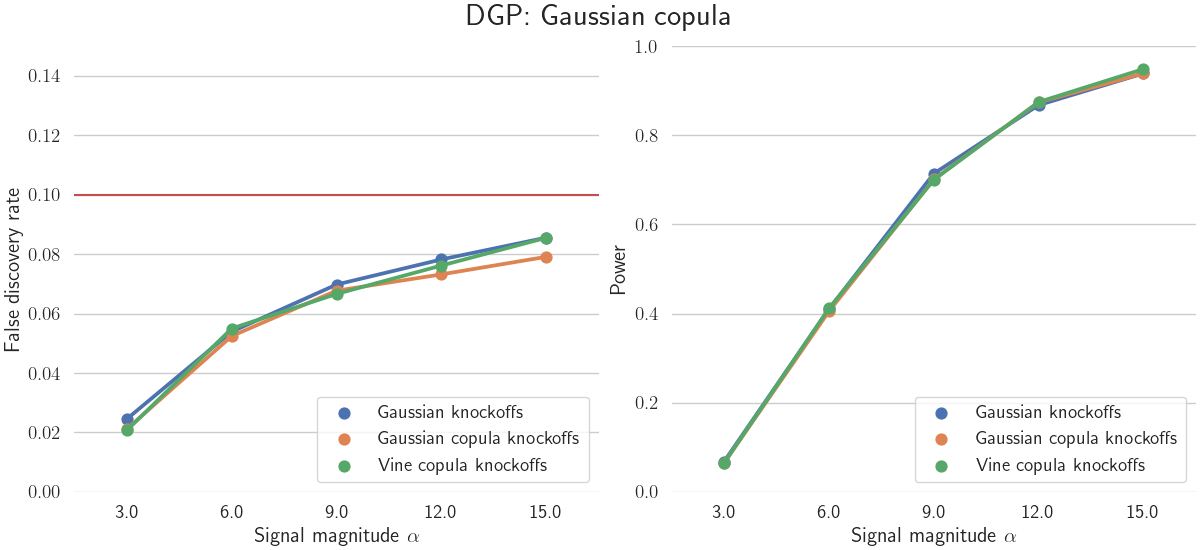}
\end{center}
\caption{
Average false discovery proportions (left) and empirical power (right) plotted against the signal magnitude $\alpha \in \lbrace 3, 6, 9, 12, 15 \rbrace$.
The DGP is the Gaussian copula \eqref{eq_gau_cop_dgp} with $d=50$ and $n=200$.
The results for Gaussian knockoffs, Gaussian copula knockoffs and vine copula knockoffs are shown color-coded and obtained with $1000$ independent repetitions.
}\label{fig_ko_study_gau_cop}
\end{figure}
The results of the controlled variable selection are shown in \autoref{fig_ko_study_gau_cop}.
In terms of the false discovery proportions and the empirical power, the three different knockoff methods perform similarly good.

\begin{figure}[!htb]
\begin{center}
\includegraphics[width=0.8\textwidth]{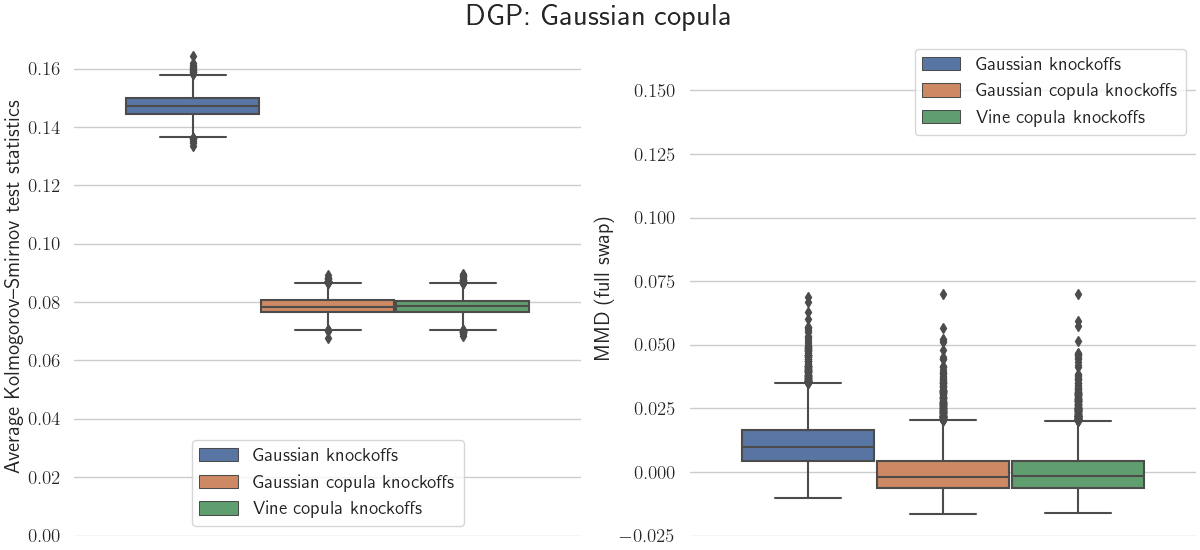}
\end{center}
\caption{
Knockoff diagnostics for the Gaussian copula \eqref{eq_gau_cop_dgp} with $d=50$ and $n=200$ as DGP.
On the left boxplots of average Kolmogorov-Smirnov test statistics for the pairs $(X_i, \widetilde{X}_i)$, $1 \leq i \leq d$ are shown.
The right plot provides boxplots of the MMD statistic for a full swap.
The results for Gaussian knockoffs, Gaussian copula knockoffs and vine copula knockoffs are shown color-coded and obtained with $1000$ independent repetitions.
}\label{fig_diagnostics_gau_cop}
\end{figure}

However, if we consider the knockoff diagnostics shown in \autoref{fig_diagnostics_gau_cop}, it becomes evident that the average Kolmogorov-Smirnov test statistics (left hand side) are lower for the Gaussian copula knockoffs and for the vine copula knockoffs in comparison to the results for the Gaussian knockoffs.
This means that the marginal distributions of the covariates $X_{1:d}$ are better replicated with the copula-based knockoff methods.
Overall this is reasonable because the univariate kernel density estimates used for the copula-based knockoff methods are by construction more flexible than simply approximating the marginal distributions with normal distributions as it is being done when using Gaussian knockoffs.
However, interestingly for the controlled variable selection it does not seem to have a major effect as can be seen in \autoref{fig_ko_study_gau_cop}.
Also in terms of the maximum mean discrepancy (MMD), shown in the boxplots on the right of \autoref{fig_diagnostics_gau_cop}, the copula-based knockoff methods are superior to the Gaussian knockoffs for the considered Gaussian copula DGP with flexible non-normal marginal distributions.

\subsection{Clayton vine copula}
The third DGP is a D-vine copula with building blocks given by (for $2\leq i \leq d$, $1 \leq j \leq i-1$)
\begin{align}
C_{j, i\ps S_{j,i}}(\cdot, \cdot) = \begin{cases}
C^{Cl}(\cdot, \cdot\ps \theta_{i-j}) & 1 \leq i-j\leq 5, \\
C^{\perp}(\cdot, \cdot) & 6 \leq i-j \leq 49.
\end{cases} \label{eq_cl_vine_dgp}
\end{align}
Here, $C^{\perp}$ denotes the independence (or product) copula and $C^{Cl}$ the Clayton copula.
The parameters of the Clayton copulas in the $(i-j)$-th tree are set to $\theta_{i-j} = \theta_1 / (1 + (i-j-1) \theta_1)$ and $\theta_1 = 2 \tau / (1-\tau)$.\footnote{
Note that the specified vine copula is the simplified vine copula representation of a d-dimensional Clayton copula, where $\tau$ is the pairwise Kendall's $\tau$.
However, for all copulas in the higher trees (from the $6$-th tree on) we use independence copulas instead of Clayton copulas, i.e., the DGP is not exactly a $d$-dimensional Clayton copula but a truncated version of its vine copula representation.}
We set $\tau=0.7$ and consider a sample size of $n=200$.
As for the Gaussian copula, we combine the Clayton vine copula \eqref{eq_cl_vine_dgp} with randomly chosen marginal distributions $F_i$, $1 \leq i \leq d$, among Gaussian mixture, Student-t and Exponential distributions (see \autoref{sec_sim_gau_cop}).
Again we obtain the random covariates via $X_i = F_i^{-1}(U_i)$, $1 \leq i \leq d$.

\begin{figure}[!htb]
\begin{center}
\includegraphics[width=0.8\textwidth]{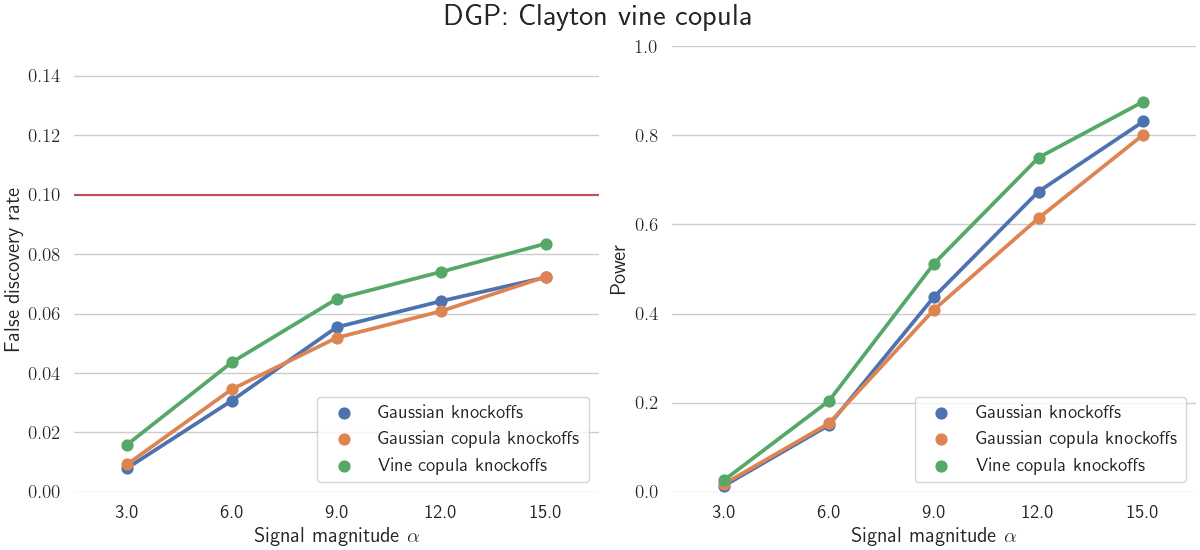}
\end{center}
\caption{
Average false discovery proportions (left) and empirical power (right) plotted against the signal magnitude $\alpha \in \lbrace 3, 6, 9, 12, 15 \rbrace$.
The DGP is the Clayton vine copula \eqref{eq_cl_vine_dgp} with $d=50$ and $n=200$.
The results for Gaussian knockoffs, Gaussian copula knockoffs and vine copula knockoffs are shown color-coded and obtained with $1000$ independent repetitions.
}\label{fig_ko_study_vine_cop}
\end{figure}

The results of the controlled variable selection are presented in \autoref{fig_ko_study_vine_cop}.
On the left of \autoref{fig_ko_study_vine_cop}, we plot the average false discovery proportion in $1000$ repetitions against the signal magnitude $\alpha$.
Comparing the false discover proportions with the nominal level $q=0.1$ shows that again with all three methods the false discoveries can be controlled.
In terms of the empirical power (right hand side of \autoref{fig_ko_study_vine_cop}) vine copula knockoffs are more powerful in comparison to the less flexible Gaussian knockoffs and Gaussian copula knockoffs.
\begin{figure}[!htb]
\begin{center}
\includegraphics[width=0.8\textwidth]{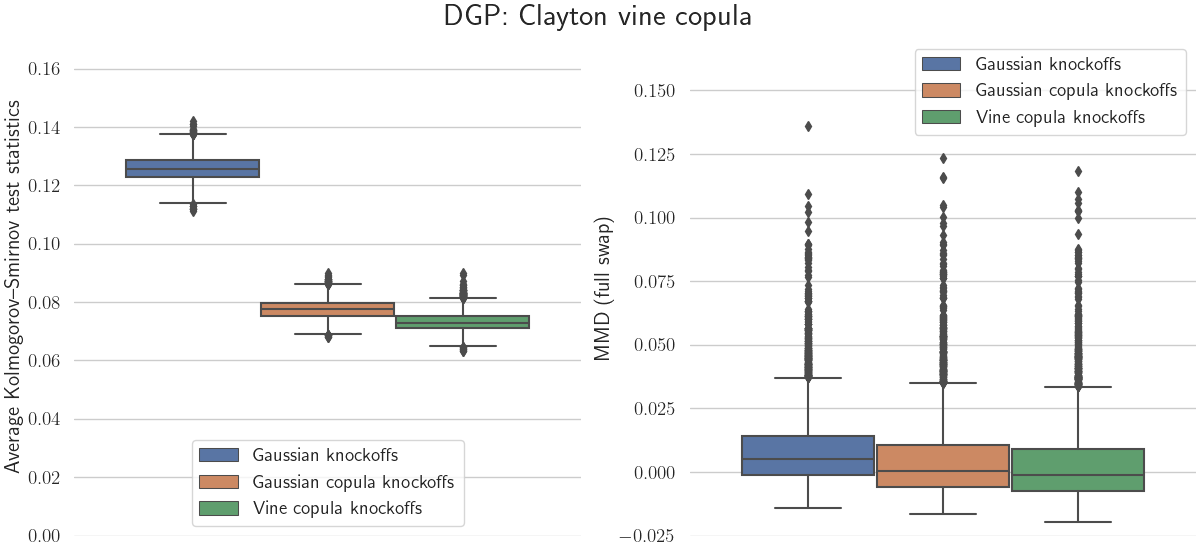}
\end{center}
\caption{
Knockoff diagnostics for the Clayton vine copula \eqref{eq_cl_vine_dgp} with $d=50$ and $n=200$ as DGP.
On the left boxplots of average Kolmogorov-Smirnov test statistics for the pairs $(X_i, \widetilde{X}_i)$, $1 \leq i \leq d$ are shown.
The right plot provides boxplots of the MMD statistic for a full swap.
The results for Gaussian knockoffs, Gaussian copula knockoffs and vine copula knockoffs are shown color-coded and obtained with $1000$ independent repetitions.
}\label{fig_diagnostics_cl_vine}
\end{figure}

The knockoffs diagnostics presented in \autoref{fig_diagnostics_cl_vine} are in line with the results of the controlled variable selection.
Vine copula knockoffs consist of flexible univariate kernel density estimates for the margins which are combined with a simplified vine copula model for modeling the dependence structure.
For the considered DGP, vine copula knockoffs are in terms of the average Kolmogorov-Smirnov test statistics and the MMD superior in comparison to the Gaussian knockoffs or Gaussian copula knockoffs.
This is reasonable as Gaussian knockoffs approximate the joint distribution with a multivariate normal distribution which might not be well suited for the considered Clayton vine copula DGP \eqref{eq_cl_vine_dgp}.
While Gaussian copula knockoffs are more flexible in terms of the univariate margins using kernel density estimates, the copula is still approximated with a Gaussian copula.
In this regard the vine copula knockoffs are more flexible and also perform better for the considered DGP as can be seen in \autoref{fig_ko_study_vine_cop} and \autoref{fig_diagnostics_cl_vine}.

%% file: outlook.tex
\section{Concluding remarks}\label{sec_concl}
In this paper we propose a vine copula based knockoff generation method for high-dimensional controlled variable selection.
We show that there is a direct connection between knockoffs and partial correlation vines.
These partial correlation vines can be naturally generalized to vine copulas.
The usage of vine copulas for generating knockoffs comes at two key advantages.
First, by using copulas, one can model the marginal distributions independently of the dependence structure and therefore benefit from highly flexible models like for example univariate kernel density estimators.
Secondly, vine copulas give access to a flexible framework to model the dependence among the covariates and their knockoff copies.
An implementation of vine copula knockoffs is provided in the \texttt{Python} package \texttt{vineknockoffs} \parencite{kurz2022}.
We further demonstrate in a simulation study that vine copula knockoff models are effective and powerful for high-dimensional controlled variable selection.

The parametric vine copula knockoff models proposed in this paper can be further tuned.
In these lines, one needs to define a loss function which balances the tradeoff between approximately satisfying the knockoff swap property and the target to render the variables and their knockoff copies as independent as possible.
For example \textcite{romano2020} use a maximum mean discrepancy (MMD) loss and \textcite{sudarshan2020} a likelihood loss to optimize deep learning models with a stochastic gradient descent algorithm.
In \autoref{sec_sgd}, we briefly discuss key components of such an approach for vine copula knockoff models.
Another interesting topic for future research is the extension of vine copula knockoffs to discrete marginal distributions, see, e.g., \textcite{panagiotelis2012,panagiotelis2017,zilko2016} for vine copula modeling with discrete or mixed discrete-continuous margins.

%% file: algos.tex
\section{Optimizing the parameters of vine copula knockoff models via a stochastic gradient descent algorithm}\label{sec_sgd}
The parametric vine copula knockoff models proposed in this paper can be further tuned.
In these lines, one needs to define a loss function which balances the tradeoff between approximately satisfying the knockoff swap property and the target to render the variables and their knockoff copies as independent as possible.
For example \textcite{romano2020} use a maximum mean discrepancy (MMD) loss and \textcite{sudarshan2020} a likelihood loss to optimize deep learning models with a stochastic gradient descent (SGD) algorithm.

Having defined a knockoff loss function, a key component for the implementation of a SGD algorithm is the implementation of the gradient of the loss with respect to the parameters.
For this, let $\theta_{k,l}$ be the parameter of the copula $C_{k, l;\, S_{k,l}}$, $2\leq l \leq 2d$, $1 \leq k \leq l-1$.
Following \autoref{algo_ko_vine}, the knockoffs $\widetilde{X}_{1:d}$ are a function of the observed variables $X_{1:d}$, the estimated marginal distributions $F_1, \ldots, F_d$ and the vine copula parameters $\theta_{k,l}$, $2\leq l \leq 2d$, $1 \leq k \leq l-1$.
Via iteratively applying the chain rule, we obtain the gradient of the knockoff loss function with respect to the parameters $\theta_{k,l}$.

In the following, we provide \autoref{algo_pits_deriv_dvine} which extends \autoref{algo_pits_dvine}.
As a result of \autoref{algo_pits_deriv_dvine} one obtains not only the PITs $W_{1:d} := \big( U_{1}, U_{2|1}, U_{3|1:2}, \ldots, U_{d|1:d-1} \big)$ but also the partial derivatives of all PITs with respect to the parameter $\theta_{k,l}$.
\begin{algorithm}[!htb]
\caption{Compute PITs and their derivative w.r.t. $\theta_{k,l}$ from a D-vine copula}
\label{algo_pits_deriv_dvine}
\DontPrintSemicolon
\KwData{$U_{1:d} \sim C_{1:d}$ and a simplified D-vine copula $\mathcal{C}_{d}^{\mathcal{V}}$}
\KwResult{$W_{1:d} := \big( U_{1}, U_{2|1}, U_{3|1:2}, \ldots, U_{d|1:d-1} \big)$
and $\widecheck{W}_{1:d} := \big( \partial_{\theta_{k,l}} U_{1}, \partial_{\theta_{k,l}} U_{2|1}, \partial_{\theta_{k,l}} U_{3|1:2}, \ldots, \partial_{\theta_{k,l}} U_{d|1:d-1} \big)$}
$W_1, a_1, b_1 \longleftarrow U_1$;
$\widecheck{W}_1, \widecheck{a}_1, \widecheck{b}_1 \longleftarrow 0$\;
\For{$i\leftarrow 2$ \KwTo $d$}{
$a_i \longleftarrow U_i$;
$\widecheck{a}_i \longleftarrow 0$;\;
\For{$j\leftarrow i-1$ \KwTo $1$}{
$d_{\theta} \longleftarrow \partial_{\theta_{k,l}} \partial_1 C_{j, i;\, S_{j,i}}(b_j, a_{j+1};\, \theta_{j,i})$\;
$d_{1} \longleftarrow \partial_1^2 C_{j, i;\, S_{j,i}}(b_j, a_{j+1};\, \theta_{j,i})$\;
$d_{2} \longleftarrow \partial_{2} \partial_1 C_{j, i;\, S_{j,i}}(b_j, a_{j+1};\, \theta_{j,i})$\;
$\widecheck{a}_{j} \longleftarrow d_{\theta} + \widecheck{b}_{j} \cdot d_1 + \widecheck{a}_{j+1} \cdot d_2$\;
$a_{j} \longleftarrow \partial_1 C_{j, i;\, S_{j,i}}(b_j, a_{j+1};\, \theta_{j,i})$\;
}
$W_i \longleftarrow a_1$;
$b_i \longleftarrow a_i$;
$\widecheck{W}_i \longleftarrow \widecheck{a}_1$;
$\widecheck{b}_i \longleftarrow \widecheck{a}_i$\;
\For{$j\leftarrow 1$ \KwTo $i-1$}{
$d_{\theta} \longleftarrow \partial_{\theta_{k,l}} \partial_2 C_{j, i;\, S_{j,i}}(b_j, a_{j+1};\, \theta_{j,i})$\;
$d_{1} \longleftarrow \partial_1 \partial_2 C_{j, i;\, S_{j,i}}(b_j, a_{j+1};\, \theta_{j,i})$\;
$d_{2} \longleftarrow \partial_2^2 C_{j, i;\, S_{j,i}}(b_j, a_{j+1};\, \theta_{j,i})$\;
$\widecheck{b}_{j} \longleftarrow d_{\theta} + \widecheck{b}_{j} \cdot d_1 + \widecheck{a}_{j+1} \cdot d_2$\;
$b_{j} \longleftarrow \partial_2 C_{j, i;\, S_{j,i}}(b_j, a_{j+1};\, \theta_{j,i})$\;
}
}
\end{algorithm}

Similarly, we provide \autoref{algo_sim_deriv_dvine} which extends \autoref{algo_sim_dvine}.
As a result of \autoref{algo_sim_deriv_dvine} one obtains not only the variables $U_{1:d} \sim C_{1:d}$ but also the partial derivatives of the variables with respect to the parameters, i.e., $\widecheck{U}_{1:d}:= \big(\partial_{\theta_{k,l}}U_1, \ldots, \partial_{\theta_{k,l}}U_d)$.
\begin{algorithm}[!htb]
\caption{Simulation from a D-vine copula with derivative w.r.t. $\theta_{k,l}$}
\label{algo_sim_deriv_dvine}
\DontPrintSemicolon
\KwData{$W_{1:d} \sim \mathcal{U}([0, 1]^d)$,
$\widecheck{W}_{1:d}:= \big(\partial_{\theta_{k,l}}W_1, \ldots, \partial_{\theta_{k,l}}W_d)$ and a simplified D-vine \mbox{copula $\mathcal{C}_{d}^{\mathcal{V}}$}}
\KwResult{$U_{1:d} \sim C_{1:d}$
and $\widecheck{U}_{1:d}:= \big(\partial_{\theta_{k,l}}U_1, \ldots, \partial_{\theta_{k,l}}U_d)$}
$U_1, a_1, b_1 \longleftarrow W_1$;
$\widecheck{U}_1, \widecheck{a}_1, \widecheck{b}_1 \longleftarrow \widecheck{W}_1$\;
\For{$i\leftarrow 2$ \KwTo $d$}{
$a_1 \longleftarrow W_i$;
$\widecheck{a}_1 \longleftarrow \widecheck{W}_i$\;
\For{$j\leftarrow 1$ \KwTo $i-1$}{
$a_{j+1} \longleftarrow \big[\partial_1 C_{j, i;\, S_{j,i}}\big]_{2}^{-1}(b_j, a_j;\, \theta_{j,i})$\;
$\widecheck{d}_{\theta} \longleftarrow \partial_{\theta_{k,l}} \partial_1 C_{j, i;\, S_{j,i}}(b_j, a_{j+1};\, \theta_{j,i})$\;
$\widecheck{d}_{1} \longleftarrow \partial_1^2 C_{j, i;\, S_{j,i}}(b_j, a_{j+1};\, \theta_{j,i})$\;
$f \longleftarrow c_{j, i;\, S_{j,i}}(b_j, a_{j+1};\, \theta_{j,i})$\;
$d_\theta \longleftarrow - \widecheck{d}_{\theta} / f$\;
$d_{1} \longleftarrow - \widecheck{d}_{1} / f$\;
$d_{2} \longleftarrow 1 / f$\;
$\widecheck{a}_{j+1} \longleftarrow d_{\theta} + \widecheck{b}_{j} \cdot d_1 + \widecheck{a}_{j+1} \cdot d_2$\;
}
$U_i, b_i \longleftarrow a_i$;
$\widecheck{U}_i, \widecheck{b}_i \longleftarrow \widecheck{a}_i$\;
\For{$j\leftarrow 1$ \KwTo $i-1$}{
$d_{\theta} \longleftarrow \partial_{\theta_{k,l}} \partial_2 C_{j, i;\, S_{j,i}}(b_j, a_{j+1};\, \theta_{j,i})$\;
$d_{1} \longleftarrow \partial_1 \partial_2 C_{j, i;\, S_{j,i}}(b_j, a_{j+1};\, \theta_{j,i})$\;
$d_{2} \longleftarrow \partial_2^2 C_{j, i;\, S_{j,i}}(b_j, a_{j+1};\, \theta_{j,i})$\;
$\widecheck{b}_{j} \longleftarrow d_{\theta} + \widecheck{b}_{j} \cdot d_1 + \widecheck{a}_{j+1} \cdot d_2$\;
$b_{j} \longleftarrow \partial_2 C_{j, i;\, S_{j,i}}(b_j, a_{j+1};\, \theta_{j,i})$\;
}
}
\end{algorithm}

Using \autoref{algo_pits_deriv_dvine} and \autoref{algo_sim_deriv_dvine}, we can now not only generate knockoffs (see \autoref{algo_ko_vine}) but also compute the gradient of every knockoff with respect to every vine copula parameter $\theta_{k,l}$, $2\leq l \leq 2d$, $1 \leq k \leq l-1$.
This allows us to further tune the vine copula knockoff parameters using a stochastic gradient descent algorithm.
A prototype implementation of such a parameter optimization for vine copula knockoff models is available in the \texttt{Python} package \texttt{vineknockoffs} \parencite{kurz2022}.
It is based on an MMD loss similarly to the one used in \textcite{romano2020}.